\documentclass[preprint,12pt]{elsarticle}

\usepackage{amssymb}

\journal{
hep-lat
}

\newcommand{\bfi}[1]{\mbox{\boldmath $#1$}}

\begin{document}

\begin{frontmatter}



\title{
Instructive discussion 
of an 
effective block algorithm 
for baryon-baryon correlators
}


\author{
Hidekatsu Nemura
}
\ead{nemura.hidekatsu.gb@u.tsukuba.ac.jp}

\address{
        Centre for Computational Sciences,
        University of Tsukuba, Tsukuba, Ibaraki, 305-8577, Japan
}

\begin{abstract}
We describe an approach for the efficient calculation of 
a large number of four-point correlation functions 
for various baryon-baryon ($BB$) channels, 
which are the primary quantities for studying the 
nuclear and hyperonic nuclear forces from lattice quantum chromodynamics. 
Using the four-point correlation function of a proton-$\Lambda$ system 
as a specific example, 
we discuss how an effective block algorithm 
significantly reduces the number of iterations. 
The effective block algorithm is applied to calculate 
52 channels of the four-point correlation functions 
from nucleon$-$nucleon to $\Xi-\Xi$, 
in order to study the complete set of 
isospin symmetric 
$BB$ interactions. 
The elapsed times measured for hybrid parallel computation 
on BlueGene/Q 
demonstrate that the performance of the present algorithm is 
reasonable for various combinations of 
the number of OpenMP threads and 
the number of MPI nodes. 
The 
numerical results are 
compared with 
the results obtained using the unified contraction algorithm 
for all computed 
sites 
of the 52 
four-point correlators. 
%
\end{abstract}

\begin{keyword}
nuclear force \sep lattice QCD \sep hyperon-nucleon interaction \sep 
hypernuclei

\end{keyword}

\end{frontmatter}


\section{Introduction\label{INTRODUCTION}}

Determining how the nuclear force is described from a fundamental 
perspective is a challenging problem in physics. 
Characterising an atomic nucleus as a nucleonic many body system
provides successful results although 
a nucleon is not a true rudimentary constituent of atomic nuclei 
but a composition of quarks and gluons 
defined in 
quantum chromodynamics (QCD), 
which is the theory of the strong interaction. 
For example, 
high-precision nucleon-nucleon ($NN$) potentials are 
available to describe the $NN$ scattering data at low energies 
as well as the deuteron properties~\cite{Machleidt:2000ge,Nagels:1978sc}. 
The energy levels of light nuclei are well reproduced by such an $NN$ 
potential together with a three-nucleon force~\cite{Pieper:2001ap,Nogga:2001cz}. 
However, 
in contrast to the normal nuclear force, 
phenomenological descriptions of 
hyperon-nucleon ($YN$) and 
hyperon-hyperon ($YY$) interactions are not 
well constrained from experimental data 
because of the short life time of hyperons. 
The precise determination of $NN$, $YN$, and $YY$ interactions 
has a large impact on the studies of both 
hypernuclei~\cite{Nemura:2002fu,Nogga:2001ef,Yamamoto:2015avw} and 
the hyperonic matter inside neutron stars~\cite{SchaffnerBielich:2008kb,Demorest2010,Antoniadis:2013pzd,Masuda:2012ed}.

Recently, a new lattice-QCD-based method for studying the inter hadronic 
interactions has been proposed~\cite{Ishii:2006ec}. 
In this method, the interhadron potential can be obtained first from 
lattice QCD by measuring the Nambu-Bethe-Salpeter (NBS) wave function. 
The observables 
such as the phase shifts and the binding energies are calculated 
using the resultant potential~\cite{Aoki:2009ji}.
This approach has been applied to various baryonic 
interactions~\cite{Murano:2011nz,Nemura:2008sp,Inoue:2010hs,Inoue:2010es,Doi:2011gq,Inoue:2011ai,HALQCD:2012aa,Inoue:2013nfe,Murano:2013xxa,Etminan:2014tya,Inoue:2014ipa,Yamada:2015cra},
and has been recently extended to systems in inelastic 
channels~\cite{Aoki:2011gt,Aoki:2012bb,Sasaki:2015ifa}. 
This approach is now called HAL QCD method because almost all the recent 
developments cited above have been provided by the HAL QCD Collaboration. 
The flavour symmetry breaking is a 
key topic 
in the study of 
the isospin symmetric baryon-baryon ($BB$) interactions 
based on the $2+1$ flavour lattice QCD. 
In such a situation, it is advantageous to calculate a 
large number of NBS wave functions of various $BB$ channels 
simultaneously in a single lattice QCD calculation. 
Therefore, an efficient approach for performing such a 
computationally demanding 
lattice QCD calculation is crucial. 

The purpose of this paper is to describe 
a practicable algorithm that can efficiently compute 
a large number of four-point correlation functions 
of various $BB$ systems. 
The contraction algorithm considered in this paper is 
different from the unified contraction algorithm~\cite{Doi:2012xd} and 
has been used 
to calculate the $\Lambda N$ and $\Sigma N$ 
potentials~\cite{Nemura:2009kc,Nemura:2010nh,Nemura:2012fm}. 
This is a reasonable approach for computing 
the various $BB$ correlators 
efficaciously. 
Methods following different approach 
for large baryon number systems are found in 
Refs.~\cite{Detmold:2012eu,Gunther:2013xj}. 
The paper 
is 
organised 
as follows:
Section~\ref{OUTLINE} outlines a basic formulation of the 
HAL QCD approach. 
Section~\ref{BLOCKALGORITHM} describes 
an approach for calculating the four-point correlation 
function by considering the $p \Lambda$ system as an example.
The present contraction algorithm is generalised to various $BB$ systems 
in Section~\ref{EXTENSION2B8B8CHANNELS}. 
In Sec.~\ref{IMPLEMENTATION} we demonstrate the 
hybrid parallel computation of the four-point correlation 
functions. 
The numerical results calculated by the hybrid parallel program 
are compared with the results from the 
unified contraction algorithm in Sec.~\ref{BENCHMARK}. 
Sec.~\ref{SUMMARY} summarises the 
paper.

\section{Outline of the HAL QCD method\label{OUTLINE}}

In the study of the nuclear force using the HAL QCD approach, 
the equal time NBS wave function with Euclidean time $t$ is 
a vital quantity, which is defined by\cite{Ishii:2006ec,Aoki:2009ji} 
\begin{equation}
 \phi_{E}(\vec{r}) {\rm e}^{-E t} = 
 \sum_{\vec{X}}
 \left\langle 0
  \left|
   B_{1,\alpha}(\vec{X}+\vec{r},t)
   B_{2,\beta}(\vec{X},t)
  \right| B=2, E
 \right\rangle,
\end{equation}
where $E=\sqrt{k^2+m_{B_{1}}^2}+\sqrt{k^2+m_{B_{2}}^2}$ is the total energy 
in the centre of mass system of a baryon number $B=2$ state with masses 
$m_{B_{1}}$ and $m_{B_{2}}$. 
$B_{1,\alpha}(x)$ ($B_{2,\beta}(x)$) denotes the local interpolating field of 
baryon $B_{1}$ ($B_{2}$). 
For simplicity, we consider a two-nucleon system 
in the isospin symmetric limit. 
Thus, $m_{B_{1}} = m_{B_{2}} = m_{N}$ and 
the $B_{1,\alpha}=p_{\alpha}$ ($B_{2,\beta}=n_{\beta}$) 
is 
the local 
interpolating field of proton (neutron) given by 
%
\begin{equation}
  p_{\alpha}(x) = \varepsilon_{abc} \left(
			 u_{a}(x) C\gamma_5 d_{b}(x)
			\right) u_{c\alpha}(x),
  \quad 
  n_{\beta}(y) = - \varepsilon_{abc} \left(
			   u_a(y) C\gamma_5 d_b(y)
			  \right) d_{c\beta}(y),
  \label{FieldOperator_N}
\end{equation}
where $u_{c\alpha}(x)$ ($d_{c\beta}(x)$) is the up (down) quark field 
with the colour indices denoted by $a,b,$ and $c$, and the Dirac spinors 
denoted by $\alpha$ and $\beta$. 
The $\varepsilon_{abc}$ is the totally anti-symmetric tensor and 
$C=\gamma_{4}\gamma_{2}$ is the charge conjugation matrix. 
For simplicity, 
we have suppressed the 
dummy 
spinor indices in the 
round brackets. 
Based on the NBS wave function, we define a non-local but energy-independent 
potential 
%
$\left(
   \frac{\nabla^2}{2\mu} - \frac{k^{2}}{2\mu}
  \right)
  \phi_{E}(\vec{r}) = 
  \int d^3r^\prime\, U(\vec{r},\vec{r^{\prime}}) 
  \phi_{E}(\vec{r^{\prime}})$ 
with the reduced mass $\mu=m_{N}/2$. 
%
An important point of the HAL QCD method is that 
the potential defined above gives the correct scattering phase shift of 
the $S$-matrix for all values of $k$ in the elastic region, 
$E < E_{\rm th} \equiv 2 m_{N} + m_{\pi}$, 
with the pion mass $m_{\pi}$, by construction. 
A more detailed account of the relation between 
the NBS wave function and the $S$-matrix in QCD is found in the 
appendix A of Ref.~\cite{Aoki:2009ji}. 

In lattice QCD calculations, 
we compute the normalised four-point correlation function defined by\cite{HALQCD:2012aa} 
\begin{eqnarray}
 &&
 {R}_{\alpha\beta}^{(J,M)}(\vec{r},t-t_0) 
 \nonumber
 \\
 &&\!\!\!\!\!\!\!\!\!\!\!\! = \!\!
 \sum_{\vec{X}}
 \!\left\langle \! 0 \!
  \left|
   B_{1,\alpha}(\vec{X}\!+\!\vec{r},t)
   B_{2,\beta}(\vec{X},t)
   \overline{{\cal J}_{B_{3} B_{4}}^{(J,M)}(t_0)}
  \right| \! 0 \!
 \right\rangle
 \! / \!
 \exp\{
 -(m_{B_1}\!+\!m_{B_2})(t\!-\!t_0)
 \},
\end{eqnarray}
where 
$\overline{{\cal J}_{B_3B_4}^{(J,M)}(t_0)}=
  \sum_{\alpha^\prime\beta^\prime}
  P_{\alpha^\prime\beta^\prime}^{(J,M)}
  \overline{B_{3,\alpha^\prime}(t_0)}
  \overline{B_{4,\beta^\prime}(t_0)}$
is a source operator that creates $B_3B_4$ (=$pn$) states with the
total angular momentum $J,M$. 
The normalised four-point function can be expressed as
\begin{eqnarray}
 &&
  {R}_{\alpha\beta}^{(J,M)}(\vec{r},t-t_0) 
  \nonumber
  \\
  &&\!\!\!\!=
   \sum_{n} A_{n}
   \sum_{\vec{X}}
   \left\langle 0
    \left|
     B_{1,\alpha}(\vec{X}+\vec{r},0)
     B_{2,\beta}(\vec{X},0)
    \right| E_{n} 
   \right\rangle
   {\rm e}^{-(E_{n}-m_{B_1}-m_{B_2})(t-t_0)}
   \nonumber
   \\
   && \qquad 
   + O({\rm e}^{-(E_{\rm th}-m_{B_{1}}-m_{B_{2}})(t-t_{0})}),
\end{eqnarray}
where $E_n$ ($|E_n\rangle$) is the eigen-energy (eigen-state)
of the six-quark system 
and 
$A_n = \sum_{\alpha^\prime\beta^\prime} P_{\alpha^\prime\beta^\prime}^{(JM)}
\langle E_n | \overline{B}_{4,\beta^\prime}
\overline{B}_{3,\alpha^\prime} | 0 \rangle$. 
At moderately large $t-t_0$ where the inelastic contribution above the 
pion production 
$O({\rm e}^{-(E_{\rm th}-2m_{N})(t-t_{0})}) = O({\rm e}^{-m_{\pi}(t-t_{0})})$ 
becomes 
exiguous, 
we can construct the non-local potential $U$ through 
$\left(
   \frac{\nabla^2}{2\mu} - \frac{k^{2}}{2\mu}
  \right)
  R(\vec{r}) = 
  \int d^3r^\prime\, U(\vec{r},\vec{r^{\prime}}) 
  R(\vec{r^{\prime}}).$ 
In lattice QCD calculations in a finite box, it is practical to use 
the velocity (derivative) expansion, 
$U(\vec{r},\vec{r^{\prime}}) = V(\vec{r},\vec{\nabla}_{r})
\delta^{3}(\vec{r} - \vec{r^{\prime}}).$ 
In the lowest few orders we have 
\begin{equation}
V(\vec{r},\vec{\nabla}_{r}) = 
\underbrace{
V_{0}(r) + V_{\sigma}(r)\vec{\sigma}_{1} \cdot \vec{\sigma}_{2} + 
V_{T}(r) S_{12}}_{V_{LO}} + 
\underbrace{
V_{LS}(r) \vec{L}\cdot (\vec{\sigma}_{1}+\vec{\sigma}_{2})}_{V_{NLO}} + 
O(\nabla^{2}),
\end{equation}
where $r=|\vec{r}|$, $\vec{\sigma}_{i}$ are the Pauli matrices acting 
on the spin space of the $i$-th baryon, 
$S_{12}=3
(\vec{r}\cdot\vec{\sigma}_{1})
(\vec{r}\cdot\vec{\sigma}_{2})/r^{2}-
\vec{\sigma}_{1}\cdot
\vec{\sigma}_{2}$ is the tensor operator, and 
$\vec{L}=\vec{r}\times (-i \vec{\nabla})$ is the angular momentum operator. 
The first three-terms constitute the leading order (LO) potential while 
the fourth term corresponds to the next-to-leading order (NLO) potential. 
By taking the non-relativistic approximation, 
$E_{n} - m_{B_{1}} - m_{B_{2}} \simeq k_{n}^{2}/(2\mu) + O(k_{n}^{4})$, 
and neglecting the $V_{\rm NLO}$ and the higher order terms, 
we obtain 
$\left(\frac{\nabla^2}{2\mu} -\frac{\partial}{\partial t}\right){ R}(\vec r,t)
\simeq 
V_{\rm LO}(\vec{r}) R(\vec r,t)$. 
%
For the spin
singlet
state, we extract the 
central potential as 
$V_C(r;J=0)=({\nabla^2\over 2\mu}-{\partial\over \partial t})
{ R}/{ R}$. 
For the spin triplet state, 
the wave function 
is decomposed into 
the $S$- 
and 
$D$-wave components as 
\begin{equation}
 \left\{
 \begin{array}{l}
  R_{\alpha\beta}(\vec{r};\ ^3S_1)={\cal P}R_{\alpha\beta}(\vec{r};J=1)
   \equiv {1\over 24} \sum_{{\cal R}\in{ O}} {\cal R}
   R_{\alpha\beta}(\vec{r};J=1),
   \\
  R_{\alpha\beta}(\vec{r};\ ^3D_1)={\cal Q}R_{\alpha\beta}(\vec{r};J=1)
   \equiv (1-{\cal P})R_{\alpha\beta}(\vec{r};J=1).
 \end{array}
 \right.
\end{equation}
Therefore, 
the Schr\"{o}dinger equation with the LO 
potentials for the spin triplet state becomes
\begin{eqnarray}
 &&
 \left\{
 \begin{array}{c}
  {\cal P} \\
  {\cal Q}
 \end{array}
 \right\}
 \times
 \left\{
  -{\nabla^2\over 2\mu} 
  +V_0(r)
  +V_\sigma(r)(\vec{\sigma}_{1}\cdot\vec{\sigma}_{2})
  +V_T(r)S_{12}
 \right\}
 { R}(\vec{r},t-t_0)
 \nonumber
 \\
 &&
 =-
 \left\{
 \begin{array}{c}
  {\cal P} \\
  {\cal Q}
 \end{array}
 \right\}
 \times
 {\partial \over \partial t}~
 { R}(\vec{r},t-t_0),
\end{eqnarray}
from which
the 
central and tensor potentials, 
$V_C(r;J=0)=V_0(r)-3V_\sigma(r)$ for $J=0$, 
$V_C(r;J=1)=V_0(r) +V_\sigma(r)$, and $V_T(r)$ for $J=1$, can be
determined\footnote{
The potential is obtained from the NBS 
wave function at moderately large imaginary time; it would be 
$t-t_{0} \gg 1/m_{\pi} \sim 1.4$~fm even for the physical pion mass. 
Furthermore, 
no single state saturation between the ground state 
and the first excited states, 
$t-t_{0} \gg (\Delta E)^{-1} = \left( (2\pi)^2/(2\mu L^2) \right)^{-1}$, 
is required for the present HAL QCD method\cite{HALQCD:2012aa}, 
which becomes 
$\left( (2\pi)^2/(2\mu L^2) \right)^{-1} \simeq 4.6$~fm 
if we consider $L\sim 6$~fm and $m_{N}\simeq 1$~GeV. 
In Ref.~\cite{Murano:2011nz}, 
the validity of the velocity expansion of 
the $NN$ potential has been examined 
in quenched lattice QCD simulations 
at $m_{\pi} \simeq 530$~MeV and $L \simeq 4.4$~fm. 
}. 

The HAL QCD method mentioned above can be applied to 
the baryon number $B=2$ systems, including strangeness 
for the $YN$ and $YY$ potentials. 
In addition to the up and down quarks, we use the strange quark operator 
$s_{c\alpha}(x)$ 
to define the interpolating operators of hyperons as 
%
\begin{equation}
  \begin{array}{l}
  \Sigma^{+}_{\alpha}(x) \!=\! - \varepsilon_{abc} \left(
				    u_a(x) C\gamma_5 s_b(x)
				   \right) u_{c\alpha}(x),
  ~
  \Sigma^{-}_{\beta}(y) \!=\! - \varepsilon_{abc} \left(
				    d_a(y) C\gamma_5 s_b(y)
				   \right) d_{c\beta}(y),
  \\
  \Sigma^{0}_{\alpha}(x) \!=\! {1\over\sqrt{2}} \left(
  X_{u,\alpha}(x) - X_{d,\alpha}(x)
  \right), 
  ~
  \Lambda_{\beta}(y) \!=\! {1\over \sqrt{6}} 
  \left(
  X_{u,\beta}(y) + X_{d,\beta}(y) - 2 X_{s,\beta}(y) 
  \right),
  \\
  \Xi^{0}_{\alpha}(x) \!=\! \varepsilon_{abc} \left(
			 u_a(x) C\gamma_5 s_b(x)
			\right) s_{c\alpha}(x),
  ~
  \Xi^{-}_{\beta}(y) \!=\! - \varepsilon_{abc} \left(
			   d_a(y) C\gamma_5 s_b(y)
			  \right) s_{c\beta}(y),
  \end{array}
  \label{FieldOperator_SLXi}
\end{equation}
where
\begin{equation}
 \begin{array}{l}
X_{u,\alpha}(x) \!=\! \varepsilon_{abc}    \left(
    d_a(x) C\gamma_5 s_b(x)
   \right) u_{c\alpha}(x), 
~
X_{d,\alpha}(x) \!=\! \varepsilon_{abc}    \left(
    s_a(x) C\gamma_5 u_b(x)
   \right) d_{c\alpha}(x),
\\
X_{s,\alpha}(x) \!=\! \varepsilon_{abc}    \left(
    u_a(x) C\gamma_5 d_b(x)
   \right) s_{c\alpha}(x).
 \end{array}
  \label{FieldOperator_xxx}
\end{equation}
%
In the flavour $SU(3)$ limit, the extension of the HAL QCD method 
to the $YN$ and $YY$ systems 
is straightforward~\cite{Inoue:2010hs,Inoue:2010es,Inoue:2011ai}. 
For the $N_{f}=2+1$ flavour lattice QCD calculations, 
the $YN$ and $YY$ potentials can be obtained in a similar fashion, 
where the mass difference between $m_{B_{1}}$ and $m_{B_{2}}$ is 
appropriately 
considered~\cite{Nemura:2008sp,Nemura:2009kc,Nemura:2010nh,Nemura:2012fm}. 
In addition, the HAL QCD method is extended to obtain the coupled-channel 
potentials above the inelastic thresholds\cite{Aoki:2011gt,Aoki:2012bb,Sasaki:2015ifa}. 

\section{The effective block algorithm\label{BLOCKALGORITHM}}

\begin{figure}[t]
  \centering \leavevmode 
  \includegraphics[width=0.264\textwidth,angle=90]{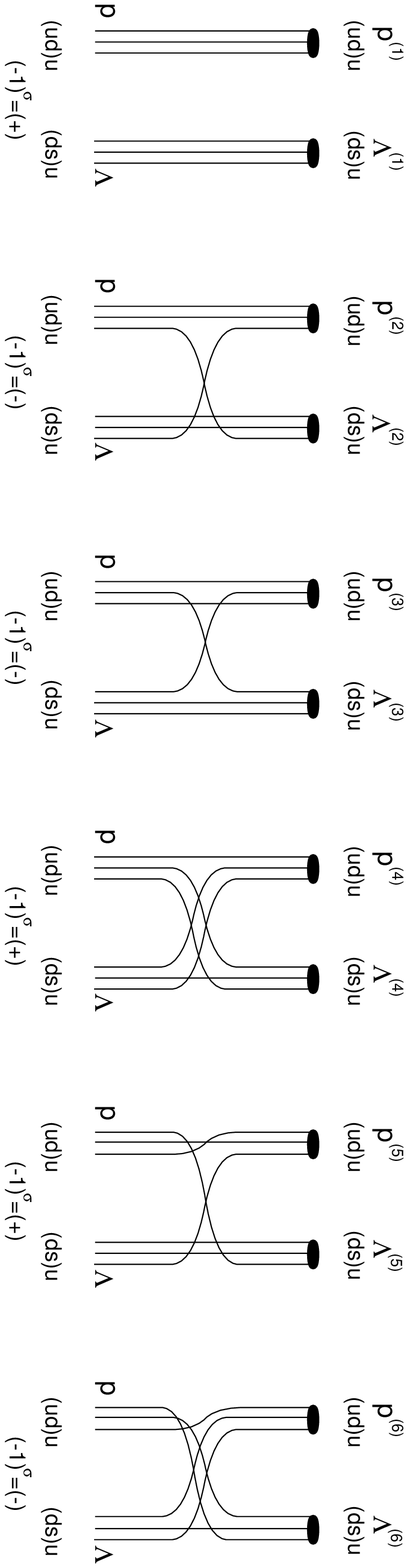}
  \footnotesize
 \caption{Diagrammatic representation of the four-point correlation function
   $\langle p\Lambda\overline{p\Lambda}\rangle$. 
   The six diagrams correspond to the six terms of Eq.~(\ref{Qexchdiag}). 
   The cyclic permutations for the quark fields 
   $(ds)u\rightarrow (su)d\rightarrow (ud)s$ 
   are taken into account in the interpolating field of $\Lambda$, 
   which correspond to the contributions from the $X_u$, $X_d$, and $X_s$. 
The parity of each permutation is also shown as $(-1)^{\sigma}$. 
 \label{Fig_FFT_Lp}}
\end{figure}

Let us consider the four-point correlation function of a 
$p\Lambda$ system as a specific example. 
In what follows, we introduce a highly abbreviated notation to indicate 
explicitly the colour, spinor, and spatial subscripts. 
For example, we 
express 
the interpolating field of proton as 
\begin{equation}
 \begin{array}{cl}
   p_\alpha(x) & = \varepsilon(c_1,c_2,c_3) (C\gamma_5)(\alpha_1,\alpha_2)
   \delta(\alpha,\alpha_3)
   u(\xi_1)d(\xi_2)u(\xi_3),  
   \qquad (\xi_i=x_i\alpha_ic_i)\\
   & =
   \varepsilon(1,2,3) (C\gamma_5)(1,2) \delta(\alpha,3)
   u(1)
   d(2)
   u(3). 
 \end{array}
\end{equation}
Here, in the last equation, the numbers 
in the 
round brackets 
show 
the indices of colour for $\varepsilon(\cdot)$, 
the indices of Dirac spinor for $(C\gamma_5)(\cdot)$ and $\delta(\cdot)$ and 
the indices both of colour, spinor, and spatial coordinate 
for the quark fields $u(\cdot)$, $d(\cdot)$, and $s(\cdot)$~\footnote{
In this paper, 
we take a 
conventional 
choice of the baryon's interpolating field given in 
Eqs.~(\ref{FieldOperator_N}), 
(\ref{FieldOperator_SLXi})$-$(\ref{FieldOperator_xxx}) which is expected 
to have large overlap with the single baryon's ground state. 
Utilising more general form of the baryon's interpolating field is 
straightforward. 
We may replace, for example, the baryon's interpolating field as 
\begin{equation}
B_{\gamma} = \varepsilon_{abc} \left( \left(
                            q_{1,a} \Gamma_1 q_{2,b}
			\right) \Gamma_2 q_{3,c} \right),
  \label{FieldOperator_B}
\end{equation}
where $q_1$, $q_2$, and $q_3$ denote particular quark flavours 
to form baryon $B$ and 
the set of gamma matrices $\{ \Gamma_1, \Gamma_2 \}$ is 
appropriately 
taken 
so as to carry 
the quantum numbers of baryon $B$ with 
combined spinor-space-time subscript $\gamma$. 
Even for the general case, we can follow the procedure in this section with 
taking two replacements everywhere: 
(i) $(C\gamma_5)(\alpha,\alpha^\prime) \rightarrow 
        \Gamma_1(\alpha,\alpha^\prime)$ and 
(ii) $\delta(\alpha,\alpha^\prime) \rightarrow 
    \Gamma_2(\alpha,\alpha^\prime)$. 
}. 
By using the abbreviated notations, 
the $p\Lambda$
four-point correlator is given by
%
\begin{eqnarray}
 &&
 {R}_{\alpha\beta
\alpha^\prime \beta^\prime
}(\vec{r},t-t_{0})
 \nonumber
 \\
 &=& 
 \sum_{\vec{X}}
 \left\langle 0
  \left|
   p_{\alpha}(\vec{X}+\vec{r},t)
   \Lambda_{\beta}(\vec{X},t)
   \overline{{\cal J}_{p_{\alpha^\prime} \Lambda_{\beta^\prime}}(t_0)}
  \right| 0 
 \right\rangle
 / 
 \exp\{
 -(m_{p}+m_{\Lambda})(t-t_0)
 \}
\nonumber
\\
&=&
  \sum_{\vec{X}}
  {1\over 6}
  {\rm e}^{(m_{p}+m_{\Lambda})(t-t_{0})}
  \varepsilon(1,4,2)
  \varepsilon(5,6,3)
  \varepsilon(1^\prime,4^\prime,2^\prime)
  \varepsilon(5^\prime,6^\prime,3^\prime)
  \nonumber
  \\
 && \times
  (C\gamma_5)(1,4) \delta(\alpha,2)
  (C\gamma_5)(1^\prime,4^\prime) \delta(\alpha^\prime,2^\prime)
  \nonumber
  \\
 &&
 \times
  \left\{
   (C\gamma_5)(5,6) \delta(\beta,3)
   + (C\gamma_5)(6,3) \delta(\beta,5)
   -2 (C\gamma_5)(3,5) \delta(\beta,6)
  \right\}
  \nonumber
  \\
 &&
 \times
  \left\{
   {
    (C\gamma_5)(5^\prime,6^\prime) \delta(\beta^\prime,3^\prime)
    + (C\gamma_5)(6^\prime,3^\prime) \delta(\beta^\prime,5^\prime)
    -2 (C\gamma_5)(3^\prime,5^\prime) \delta(\beta^\prime,6^\prime)
   }
  \right\}
 \nonumber
 \\
 &&
 \times
  \langle
   u(1) d(4) u(2)
   d(5) s(6) u(3)
   \bar{u}(3^\prime) \bar{s}(6^\prime) \bar{d}(5^\prime)
   \bar{u}(2^\prime) \bar{d}(4^\prime) \bar{u}(1^\prime)
  \rangle,
  \label{LN.pLLp.ududsu}
\end{eqnarray}
where
\begin{equation}
\vec{x}_1 = \vec{x}_2 = \vec{x}_4 = \vec{X} + \vec{r}, 
\qquad 
\vec{x}_3 = \vec{x}_5 = \vec{x}_6 = \vec{X}. 
\end{equation}
The last line in Eq.~(\ref{LN.pLLp.ududsu}) 
is 
evaluated through the Wick's contraction and 
represented in terms of quark propagators 
$\langle q(\xi_{i}) \overline{q}(\xi^\prime_{j})\rangle = \langle q({i}) \overline{q}({j}^\prime)\rangle$, 
\begin{eqnarray}
 &&
  \langle
   u(1) d(4) u(2)
   d(5) s(6) u(3)
   \bar{u}(3^\prime) \bar{s}(6^\prime) \bar{d}(5^\prime)
   \bar{u}(2^\prime) \bar{d}(4^\prime) \bar{u}(1^\prime)
  \rangle
  \nonumber
  \\
 &=&
  \left(
  {
\langle u(3) \bar{u}(3^\prime) \rangle }
  \det\left|
       \begin{array}{cc}
        \langle u(1) \bar{u}(1^\prime) \rangle &
	\langle u(1) \bar{u}(2^\prime) \rangle 
	\\
        \langle u(2) \bar{u}(1^\prime) \rangle &
	\langle u(2) \bar{u}(2^\prime) \rangle 
       \end{array}
      \right|
   \langle d(4) \bar{d}(4^\prime) \rangle
   \langle d(5) \bar{d}(5^\prime) \rangle
   \right.
   \nonumber
   \\
 &&
   -
   {
\langle u(3) \bar{u}(2^\prime) \rangle }
   \det\left|
	\begin{array}{cc}
	 \langle u(1) \bar{u}(1^\prime) \rangle &
	 \langle u(1) \bar{u}(3^\prime) \rangle
	 \\
	 \langle u(2) \bar{u}(1^\prime) \rangle &
	  \langle u(2) \bar{u}(3^\prime) \rangle
	\end{array}
       \right|
   \langle d(4) \bar{d}(4^\prime) \rangle
   \langle d(5) \bar{d}(5^\prime) \rangle
   \nonumber
   \\
 &&
  -
  {
\langle u(3) \bar{u}(3^\prime) \rangle }
  \det\left|
       \begin{array}{cc}
        \langle u(1) \bar{u}(1^\prime) \rangle &
	\langle u(1) \bar{u}(2^\prime) \rangle 
	\\
        \langle u(2) \bar{u}(1^\prime) \rangle &
	\langle u(2) \bar{u}(2^\prime) \rangle 
       \end{array}
      \right|
   \langle d(4) \bar{d}(5^\prime) \rangle
   \langle d(5) \bar{d}(4^\prime) \rangle
   \nonumber
   \\
 &&
   +
   {
\langle u(3) \bar{u}(2^\prime) \rangle }
   \det\left|
	\begin{array}{cc}
	 \langle u(1) \bar{u}(1^\prime) \rangle &
	 \langle u(1) \bar{u}(3^\prime) \rangle
	 \\
	 \langle u(2) \bar{u}(1^\prime) \rangle &
	  \langle u(2) \bar{u}(3^\prime) \rangle
	\end{array}
       \right|
   \langle d(4) \bar{d}(5^\prime) \rangle
   \langle d(5) \bar{d}(4^\prime) \rangle
   \nonumber
   \\
 &&
   +
   {
\langle u(3) \bar{u}(1^\prime) \rangle}
   \det\left|
	\begin{array}{cc}
	 \langle u(1) \bar{u}(2^\prime) \rangle &
	 \langle u(1) \bar{u}(3^\prime) \rangle
	 \\
	 \langle u(2) \bar{u}(2^\prime) \rangle &
	 \langle u(2) \bar{u}(3^\prime) \rangle
	\end{array}
       \right|
   \langle d(4) \bar{d}(4^\prime) \rangle
   \langle d(5) \bar{d}(5^\prime) \rangle
   \nonumber
   \\
 &&
   \left.
   -
   {
\langle u(3) \bar{u}(1^\prime) \rangle}
   \det\left|
	\begin{array}{cc}
	 \langle u(1) \bar{u}(2^\prime) \rangle &
	 \langle u(1) \bar{u}(3^\prime) \rangle
	 \\
	 \langle u(2) \bar{u}(2^\prime) \rangle &
	 \langle u(2) \bar{u}(3^\prime) \rangle
	\end{array}
       \right|
   \langle d(4) \bar{d}(5^\prime) \rangle
   \langle d(5) \bar{d}(4^\prime) \rangle
   \right)
   \nonumber
   \\
 &&
   \times 
   \langle s(6) \bar{s}(6^\prime) \rangle. 
\label{Qexchdiag}
\end{eqnarray}
The six terms in Eq.~(\ref{Qexchdiag}) can be depicted 
with six diagrams as shown in Fig.~\ref{Fig_FFT_Lp}. 
%
%

%
The Eq.~(\ref{LN.pLLp.ududsu}) includes implicit summations such as 
$\sum_{c_1,\cdots,c_6}$ $\sum_{\alpha_1,\cdots,\alpha_6}$ 
$\sum_{c_1^\prime,\cdots,c_6^\prime}$ $\sum_{\alpha_1^\prime,\cdots,\alpha_6^\prime}$;
the number of iterations for each summation is 
$N_c=3$ for the colour or $N_\alpha=4$ for the Dirac spinor. 
Combining the iteration due to the Wick contraction,
for the system with the baryon number $B$ in general, 
the total number of iterations for such a correlator 
is in a naive counting 
$(N_c!N_\alpha)^{2B} \times N_{u}!N_{d}!N_{s}!$, 
where the 
$N_{u}, N_{d}$, and $N_{s}$ are the numbers of 
$u$-quark, $d$-quark, and $s$-quark, respectively; 
thus the numbers satisfy $N_{u} + N_{d} + N_{s} = 3B$.
Clearly, the above 
counting is too naive though 
curtailment of the number of iterations 
is not trivial. 
We now explain briefly how the number of iterations reduces when 
we calculate the four-point correlation function of the $p\Lambda$ 
system~\cite{Nemura:2009kc}. 
In Ref.~\cite{Nemura:2008sp}, only the limited spatial points were 
evaluated on a $L^3$ lattice because of the computational cost 
$O(L^6)$ in the primitive numerical approach. 
In this paper we employ the Fast-Fourier-Transform (FFT) to improve the 
numerical performance to $O(L^3 \log L^3)$; 
we consider the diagrammatic classification of the Wick contraction 
in order to make better use of the FFT. 
%
\begin{eqnarray}
 {R}_{\alpha\beta
\alpha^\prime \beta^\prime
}(\vec{r})
 &=&
  \sum_{i=1}^{6} F_i
  \sum_{\vec{X}} 
  \left(
       [p^{(i)}_\alpha](\vec{X}+\vec{r}) \times
       [\Lambda^{(i)}_\beta](\vec{X})
  \right)_{\alpha^\prime \beta^\prime}
  \nonumber
  \\
 &=&
  {1\over L^3}
  \sum_{\vec{q}}
  \left(
   \sum_{i=1}^{6} F_i
   \left(
        [\widetilde{p^{(i)}_\alpha}](\vec{q}) \times 
        [\widetilde{\Lambda^{(i)}_\beta}](-\vec{q})
   \right)_{\alpha^\prime \beta^\prime}
 \right)
  {\rm e}^{i\vec{q}\cdot\vec{r}},
  \label{LN.pLLp.FFT}
\end{eqnarray}
where 
$[\widetilde{p^{(i)}_\alpha}](\vec{q}) = 
\sum_{\vec{x}}[p^{(i)}_\alpha](\vec{x}) {\rm e}^{-i\vec{q}\cdot\vec{x}}$, 
$[\widetilde{\Lambda^{(i)}_\beta}](\vec{q}) = 
\sum_{\vec{x}}[\Lambda^{(i)}_\beta](\vec{x}) {\rm e}^{-i\vec{q}\cdot\vec{x}}$, 
and
$F_i=(-1)^{\sigma_i}
(1/6)
{\rm e}^{(m_{p}+m_{\Lambda})(t-t_0)}$ 
with $\sigma_i={\rm even (odd)}$ for the even (odd) permutations. 
We omit the explicit $(t-t_{0})$ dependence both of 
$R_{\alpha\beta\alpha^\prime\beta^\prime}$, 
$[p_{\alpha}^{(i)}]$, and $[\Lambda_{\beta}^{(i)}]$. 
Six diagrams in Fig.~\ref{Fig_FFT_Lp} correspond to 
the six baryon-block pairs 
$([{p^{(1)}_\alpha}] \times [{\Lambda^{(1)}_\beta}]), \cdots, 
([{p^{(6)}_\alpha}] \times [{\Lambda^{(6)}_\beta}])$. 
Note that the number of diagrams is reduced by the 
factor $2^{B-N_\Lambda-N_{\Sigma^0}}$ 
since the exchange between identical quarks in each baryon operator 
in the sink 
shall be taken into account in the construction of each baryon block
$[{p^{(i)}_\alpha}]$ or 
$[{\Lambda^{(i)}_\beta}]$, where $N_\Lambda (N_{\Sigma^0})$ is the 
number of $\Lambda$ ($\Sigma^0$) in the sink. 
We present the 
explicit forms of the baryon blocks. 
For simplicity, 
we consider only the contributions from $\overline{X}_{u}$ in the 
$\overline{\Lambda}_{\beta^\prime}$ in the source for a while 
and 
omit the contributions from the $\overline{X}_{d}$ and 
$\overline{X}_{s}$ operators.  
The contributions from $\overline{X}_d$ and $\overline{X}_s$ are 
discussed later. 

\bigskip \noindent
%
(i)~{\bf $\bfi{p_{\alpha}^{(1)}}$ and $\bfi{\Lambda_{\beta}^{(1)}}$}

The first diagram is the simplest case:
\begin{equation}
 {R}_{\alpha\beta\alpha^\prime\beta^\prime}^{(1)}(\vec{r})
 = \sum_{\vec{X}} \left([p_{\alpha}^{(1)}](\vec{X}+\vec{r}) \times
  [\Lambda_{\beta}^{(1)}](\vec{X})\right)_{\alpha^\prime\beta^\prime}
 = \sum_{\vec{X}} [p_{\alpha\alpha^\prime}^{(1)}](\vec{X}+\vec{r})
  [\Lambda_{\beta\beta^\prime}^{(1)}](\vec{X}), 
\label{convolution1_pLLp}
\end{equation}
where 
\begin{eqnarray}
 &&
  [p_{\alpha\alpha^\prime}^{(1)}](\vec{x})
  \nonumber
  \\
 &=& 
  \varepsilon(1,4,2) (C\gamma_5)(1,4) \delta(\alpha,2)
  \varepsilon(1^\prime,4^\prime,2^\prime)
  (C\gamma_5)(1^\prime,4^\prime) \delta(\alpha^\prime,2^\prime)
  \nonumber
  \\
 &&
  \times 
  \det\left|
       \begin{array}{cc}
        \langle u(1) \bar{u}({1^\prime}) \rangle &
	\langle u(1) \bar{u}(2^\prime) \rangle 
	\\
        \langle u(2) \bar{u}(1^\prime) \rangle &
	\langle u(2) \bar{u}({2^\prime}) \rangle 
       \end{array}
      \right|
  \langle d(4) \bar{d}({4^\prime}) \rangle, 
\label{baryonblock_p1_pLLp}
  \\
 &&
  [\Lambda_{\beta\beta^\prime}^{(1)}](\vec{y})
  \nonumber
  \\
 &=&
  \varepsilon(5,6,3)
  \left\{
   (C\gamma_5)(5,6) \delta(\beta,3)
   + (C\gamma_5)(6,3) \delta(\beta,5)
   -2 (C\gamma_5)(3,5) \delta(\beta,6)
  \right\}
  \nonumber
  \\
 &&
  \times 
  \varepsilon(5^\prime,6^\prime,3^\prime)
             {(C\gamma_5)(5^\prime,6^\prime) \delta(\beta^\prime,3^\prime)}
      {\langle u(3) \bar{u}({3^\prime}) \rangle} 
      \langle d(5) \bar{d}({5^\prime}) \rangle
      \langle s(6) \bar{s}({6^\prime}) \rangle. 
\end{eqnarray}
This is just a product of two two-point 
correlation functions. 
The summations of all internal indices can be performed prior 
to evaluating the FFT. 
This fact significantly slashes in the computational cost; 
the reduction factor at the first diagram is 
$(N_c!N_\alpha)^2\times2^{B-N_\Lambda-N_{\Sigma^0}}/1 = 1152$. 

\bigskip \noindent
%
(ii)~{\bf {$\bfi{p_{\alpha}^{(2)}}$ and $\bfi{\Lambda_{\beta}^{(2)}}$}}

The second diagram 
shows 
an one-quark exchange 
in $u$ quarks: 
\begin{eqnarray}
 {R}_{\alpha\beta\alpha^\prime\beta^\prime}^{(2)}(\vec{r})
 &=& \sum_{\vec{X}}
 \left(
  [p_{\alpha}^{(2)}](\vec{X}+\vec{r})
  \times
  [\Lambda_{\beta}^{(2)}](\vec{X})
 \right)_{\alpha^\prime\beta^\prime}
  \nonumber
  \\
 &=& 
  \sum_{\vec{X}}
  \sum_{c_2^\prime,c_3^\prime
}
  [p_{\alpha\beta^\prime}^{(2)}](\vec{X}+\vec{r};
  {
c_2^\prime,c_3^\prime
}
  )
  [\Lambda_{\beta\alpha^\prime}^{(2)}](\vec{X};
  {
c_2^\prime,c_3^\prime
}
  ), 
\end{eqnarray}
where
\begin{eqnarray}
 &&
  [p_{\alpha\beta^\prime}^{(2)}]
  (\vec{x};c_2^\prime,c_3^\prime)
  \nonumber
  \\
 &=&
  \varepsilon(1,4,2) (C\gamma_5)(1,4) \delta(\alpha,2)
  \varepsilon(1^\prime,4^\prime,2^\prime)
  (C\gamma_5)(1^\prime,4^\prime) 
  \delta(\beta^\prime,3^\prime)
  \nonumber
  \\
 &&
  \times
  \det\left|
        \begin{array}{cc}
	 \langle u(1) \bar{u}({1^\prime}) \rangle &
	 \langle u(1) \bar{u}(3^\prime) \rangle
	 \\
	 \langle u(2) \bar{u}(1^\prime) \rangle &
	  \langle u(2) \bar{u}({3^\prime}) \rangle
	\end{array}
      \right|
  \langle d(4) \bar{d}({4^\prime}) \rangle, 
  \\
 &&
  [\Lambda_{\beta\alpha^\prime}^{(2)}]
  (\vec{y};c_2^\prime,c_3^\prime)
  \nonumber
  \\
 &=&
  \varepsilon(5,6,3)
  \left\{
   (C\gamma_5)(5,6) \delta(\beta,3)
   + (C\gamma_5)(6,3) \delta(\beta,5)
   -2 (C\gamma_5)(3,5) \delta(\beta,6)
  \right\}
  \nonumber
  \\
 &&
  \times
  \varepsilon(5^\prime,6^\prime,3^\prime)
             {(C\gamma_5)(5^\prime,6^\prime)}
  \delta(\alpha^\prime,2^\prime)
      {\langle u(3) \bar{u}({2^\prime}) \rangle}
      \langle d(5) \bar{d}({5^\prime}) \rangle
      \langle s(6) \bar{s}({6^\prime}) \rangle. 
\end{eqnarray}
We have additional arguments, 
($c_2^\prime,c_3^\prime$),
for the baryon blocks $[p_{\alpha}^{(2)}]$ and $[\Lambda_{\beta}^{(2)}]$ 
because of 
the exchange of the quark fields 
in the source. 
Note that 
the $\delta(\alpha^\prime,2^\prime)$ in $\overline{p}_{\alpha^\prime}$ and 
the $\delta(\beta^\prime,3^\prime)$ in $\overline{\Lambda}_{\beta^\prime}$ are
also 
exchanged between the baryon 
blocks $[p_{\alpha\beta^\prime}^{(2)}]$ and 
$[\Lambda_{\beta\alpha^\prime}^{(2)}]$ so that the 
two outer indices in the source $(\alpha^\prime\beta^\prime)$ 
are crossed as 
$[p_{\alpha\beta^\prime}^{(2)}]$ and 
$[\Lambda_{\beta\alpha^\prime}^{(2)}]$. 
Performed 
these manipulations, 
the number of explicit summations of indices reduces to 
only 
two colours 
which makes the 
reduction factor 
$(N_c!N_\alpha)^2\times2^{B-N_\Lambda-N_{\Sigma^0}}/(N_c^2)= 128$. 
\bigskip \noindent
%
(iii)~{\bf {$\bfi{p_{\alpha}^{(3)}}$ and $\bfi{\Lambda_{\beta}^{(3)}}$}}

This case has an exchange in $d$ quarks: 
\begin{eqnarray}
 {R}_{\alpha\beta\alpha^\prime\beta^\prime}^{(3)}(\vec{r})
 &\!\!\!\!=&\!\!\!\! \sum_{\vec{X}}
 \left(
  [p_{\alpha}^{(3)}](\vec{X}+\vec{r})
  \times
  [\Lambda_{\beta}^{(3)}](\vec{X})
 \right)_{\alpha^\prime\beta^\prime}
  \nonumber
  \\
 &\!\!\!\!=&\!\!\!\! \sum_{\vec{X}}
  \!\!\sum_{c_4^\prime,c_5^\prime,\alpha_4^\prime,\alpha_5^\prime}\!\!\!\!\!\!
  [p_{\alpha\alpha^\prime}^{(3)}](\vec{X}\!\!+\!\vec{r};
  {
c_4^\prime,c_5^\prime,\alpha_4^\prime,\alpha_5^\prime}
  )
  [\Lambda_{\beta\beta^\prime}^{(3)}](\vec{X};
  {
c_4^\prime,c_5^\prime,\alpha_4^\prime,\alpha_5^\prime}
  ), 
\end{eqnarray}
where
\begin{eqnarray}
 &&
 [p_{\alpha\alpha^\prime}^{(3)}]
 (\vec{x};c_4^\prime,c_5^\prime,\alpha_4^\prime,\alpha_5^\prime)
 \nonumber
 \\
 &=&
  \varepsilon(1,4,2) (C\gamma_5)(1,4) \delta(\alpha,2)
  \varepsilon(1^\prime,4^\prime,2^\prime)
  (C\gamma_5)(1^\prime,4^\prime) \delta(\alpha^\prime,2^\prime)
 \nonumber
  \\
 && \qquad \times 
  \det\left|
       \begin{array}{cc}
        \langle u(1) \bar{u}({1^\prime}) \rangle &
	\langle u(1) \bar{u}(2^\prime) \rangle 
	\\
        \langle u(2) \bar{u}(1^\prime) \rangle &
	\langle u(2) \bar{u}({2^\prime}) \rangle 
       \end{array}
      \right|
  \langle d(4) \bar{d}({5^\prime}) \rangle, 
  \\
 &&
 [\Lambda_{\beta\beta^\prime}^{(3)}]
 (\vec{y};c_4^\prime,c_5^\prime,\alpha_4^\prime,\alpha_5^\prime)
 \nonumber
 \\
 &=&
  \varepsilon(5,6,3)
  \left\{
   (C\gamma_5)(5,6) \delta(\beta,3)
   + (C\gamma_5)(6,3) \delta(\beta,5)
   -2 (C\gamma_5)(3,5) \delta(\beta,6)
  \right\}
 \nonumber
  \\
 && \times 
  \varepsilon(5^\prime,6^\prime,3^\prime)
             {(C\gamma_5)(5^\prime,6^\prime) \delta(\beta^\prime,3^\prime)}
  {\langle u(3) \bar{u}({3^\prime}) \rangle}
  \langle d(5) \bar{d}({4^\prime}) \rangle
  \langle s(6) \bar{s}({6^\prime}) \rangle. 
\end{eqnarray}
The number of explicit summations of indices reduces to 
two colours and two spinors, which makes the 
reduction factor 
$(N_c!N_\alpha)^2\times2^{B-N_\Lambda-N_{\Sigma^0}}/(N_c^2 N_\alpha^2)= 8$.

\bigskip \noindent
%
(iv)~{\bf {$\bfi{p_{\alpha}^{(4)}}$ and $\bfi{\Lambda_{\beta}^{(4)}}$}}

This is one of two-quark exchange diagrams 
in the $\langle p\Lambda\overline{p\Lambda}\rangle$: 
\begin{eqnarray}
 {R}_{\alpha\beta\alpha^\prime\beta^\prime}^{(4)}(\vec{r})
 &\!\!\!\!=&\!\!\!\! \sum_{\vec{X}}
 \left(
  [p_{\alpha}^{(4)}](\vec{X}+\vec{r})
  \times
  [\Lambda_{\beta}^{(4)}](\vec{X})
 \right)_{\alpha^\prime\beta^\prime}
  \nonumber
  \\
 &\!\!\!\!=&\!\!\!\! \sum_{\vec{X}}
  \!\!\sum_{c_1^\prime,c_6^\prime,\alpha_1^\prime,\alpha_6^\prime}\!\!\!\!
  [p_{\alpha\beta^\prime}^{(4)}](\vec{X}\!\!+\!\vec{r};
  {
c_1^\prime,c_6^\prime,\alpha_1^\prime,\alpha_6^\prime}
  )
  [\Lambda_{\beta\alpha^\prime}^{(4)}](\vec{X};
  {
c_1^\prime,c_6^\prime,\alpha_1^\prime,\alpha_6^\prime}
  ), 
\end{eqnarray}
where
\begin{eqnarray}
 &&
  [p_{\alpha\beta^\prime}^{(4)}]
  (\vec{x};c_1^\prime,c_6^\prime,\alpha_1^\prime,\alpha_6^\prime)
 \nonumber
  \\
 &=&
  \varepsilon(1,4,2) (C\gamma_5)(1,4) \delta(\alpha,2)
  \varepsilon(5^\prime,6^\prime,3^\prime)
             {(C\gamma_5)(5^\prime,6^\prime) \delta(\beta^\prime,3^\prime)}
  \nonumber 
  \\
 && \qquad \times
  \det\left|
        \begin{array}{cc}
	 \langle u(1) \bar{u}({1^\prime}) \rangle &
	 \langle u(1) \bar{u}(3^\prime) \rangle
	 \\
	 \langle u(2) \bar{u}(1^\prime) \rangle &
	  \langle u(2) \bar{u}({3^\prime}) \rangle
	\end{array}
      \right|
  \langle d(4) \bar{d}({5^\prime}) \rangle, 
  \\
 &&
 [\Lambda_{\beta\alpha^\prime}^{(4)}]
 (\vec{y};c_1^\prime,c_6^\prime,\alpha_1^\prime,\alpha_6^\prime)
 \nonumber
 \\
 &=&
  \varepsilon(5,6,3)
  \left\{
   (C\gamma_5)(5,6) \delta(\beta,3)
   + (C\gamma_5)(6,3) \delta(\beta,5)
   -2 (C\gamma_5)(3,5) \delta(\beta,6)
  \right\}
 \nonumber
  \\
 && \times
  \varepsilon(1^\prime,4^\prime,2^\prime)
  (C\gamma_5)(1^\prime,4^\prime) \delta(\alpha^\prime,2^\prime)
  {\langle u(3) \bar{u}({2^\prime}) \rangle}
  \langle d(5) \bar{d}({4^\prime}) \rangle
  \langle s(6) \bar{s}({6^\prime}) \rangle. 
\end{eqnarray}
Note that 
two tensorial factors 
$\varepsilon(5^\prime,6^\prime,3^\prime)
(C\gamma_5)(5^\prime,6^\prime) \delta(\beta^\prime,3^\prime)$
and 
\\
$\varepsilon(1^\prime,4^\prime,2^\prime)
(C\gamma_5)(1^\prime,4^\prime) \delta(\alpha^\prime,2^\prime)$
are exchanged between $[p_{\alpha\beta^\prime}^{(4)}]$ 
and $[\Lambda_{\beta\alpha^\prime}^{(4)}]$ due to 
the two-quark exchange 
so that the two outer source indices $(\alpha^\prime, \beta^\prime)$ 
are exchanged, too. 
The number of explicit summations of indices reduces to 
two colours and two spinors, which makes the 
reduction factor 
$(N_c!N_\alpha)^2\times2^{B-N_\Lambda-N_{\Sigma^0}}/(N_c^2 N_\alpha^2)= 8$.

\bigskip \noindent
%
(v)~{\bf {$\bfi{p_{\alpha}^{(5)}}$ and $\bfi{\Lambda_{\beta}^{(5)}}$}}

In this case we have another exchange diagram in $u$ quarks: 
\begin{eqnarray}
 {R}_{\alpha\beta\alpha^\prime\beta^\prime}^{(5)}(\vec{r})
 &=& \sum_{\vec{X}}
 \left(
  [p_{\alpha}^{(5)}](\vec{X}+\vec{r})
  \times
  [\Lambda_{\beta}^{(5)}](\vec{X})
 \right)_{\alpha^\prime\beta^\prime}
  \nonumber
  \\
 &=& \sum_{\vec{X}}
  \sum_{c_1^\prime,c_3^\prime,\alpha_1^\prime
}
  [p_{\alpha\alpha^\prime\beta^\prime}^{(5)}](\vec{X}+\vec{r};
  {
c_1^\prime,c_3^\prime,\alpha_1^\prime
}
  )
  [\Lambda_{\beta}^{(5)}](\vec{X};
  {
c_1^\prime,c_3^\prime,\alpha_1^\prime
}
  ), 
\end{eqnarray}
where
\begin{eqnarray}
 &&
 [p_{\alpha\alpha^\prime\beta^\prime}^{(5)}]
  (\vec{x};c_1^\prime,c_3^\prime,\alpha_1^\prime)
  \nonumber
  \\
 &=&
  \varepsilon(1,4,2) (C\gamma_5)(1,4) \delta(\alpha,2)
  \varepsilon(1^\prime,4^\prime,2^\prime)
  (C\gamma_5)(1^\prime,4^\prime) \delta(\alpha^\prime,2^\prime)
  \delta(\beta^\prime,3^\prime)
  \nonumber
  \\
 && \qquad \times
  \det\left|
        \begin{array}{cc}
	 \langle u(1) \bar{u}({2^\prime}) \rangle &
	 \langle u(1) \bar{u}(3^\prime) \rangle
	 \\
	 \langle u(2) \bar{u}(2^\prime) \rangle &
	 \langle u(2) \bar{u}({3^\prime}) \rangle
	\end{array}
      \right|
  \langle d(4) \bar{d}({4^\prime}) \rangle, 
  \\
 &&
 [\Lambda_{\beta}^{(5)}]
 (\vec{y};c_1^\prime,c_3^\prime,\alpha_1^\prime)
 \nonumber
 \\
 &=&
  \varepsilon(5,6,3)
  \left\{
   (C\gamma_5)(5,6) \delta(\beta,3)
   + (C\gamma_5)(6,3) \delta(\beta,5)
   -2 (C\gamma_5)(3,5) \delta(\beta,6)
  \right\}
  \nonumber
  \\
 && \times 
  \varepsilon(5^\prime,6^\prime,3^\prime)
             {(C\gamma_5)(5^\prime,6^\prime)}
  {\langle u(3) \bar{u}({1^\prime}) \rangle}
  \langle d(5) \bar{d}({5^\prime}) \rangle
  \langle s(6) \bar{s}({6^\prime}) \rangle. 
\end{eqnarray}
Note that both 
the $\delta(\beta^\prime,3^\prime)$ in $\overline{\Lambda}_{\beta^\prime}$ 
and the $\delta(\alpha^\prime,2^\prime)$ in $\overline{p}_{\alpha^\prime}$ 
transfer to the baryon 
block $[p_{\alpha\alpha^\prime\beta^\prime}^{(5)}]$ so that the 
two outer indices in the source $(\alpha^\prime\beta^\prime)$ are 
accompanied in the $[p_{\alpha\alpha^\prime\beta^\prime}^{(5)}]$. 
The number of explicit summations of indices reduces to 
two colours and one spinor, which makes the 
reduction factor 
$(N_c!N_\alpha)^2\times2^{B-N_\Lambda-N_{\Sigma^0}}/(N_c^2 N_\alpha)= 32$.

\bigskip \noindent
%
(vi)~{\bf {$\bfi{p_{\alpha}^{(6)}}$ and $\bfi{\Lambda_{\beta}^{(6)}}$}}

In this case we have another two-quark exchange diagram: 
\begin{eqnarray}
 {R}_{\alpha\beta\alpha^\prime\beta^\prime}^{(6)}(\vec{r})
 &=& \sum_{\vec{X}}
 \left(
  [p_{\alpha}^{(6)}](\vec{X}+\vec{r})
  \times
  [\Lambda_{\beta}^{(6)}](\vec{X})
 \right)_{\alpha^\prime\beta^\prime}
  \nonumber
  \\
 &=& \sum_{\vec{X}}
  \sum_{c_2^\prime,c_6^\prime,
\alpha_6^\prime}
  [p_{\alpha\alpha^\prime\beta^\prime}^{(6)}](\vec{X}+\vec{r};
  {
c_2^\prime,c_6^\prime,
\alpha_6^\prime}
  )
  [\Lambda_{\beta}^{(6)}](\vec{X};
  {
c_2^\prime,c_6^\prime,
\alpha_6^\prime}
  ), 
\end{eqnarray}
where
\begin{eqnarray}
 &&
 [p_{\alpha\alpha^\prime\beta^\prime}^{(6)}]
  (\vec{x};c_2^\prime,c_6^\prime,\alpha_6^\prime)
  \nonumber
  \\
 &=&
  \varepsilon(1,4,2) (C\gamma_5)(1,4) \delta(\alpha,2)
             {\delta(\alpha^\prime,2^\prime)}
  \varepsilon(5^\prime,6^\prime,3^\prime)
             {(C\gamma_5)(5^\prime,6^\prime) \delta(\beta^\prime,3^\prime)}
  \nonumber 
  \\
 && \qquad \times
  \det\left|
        \begin{array}{cc}
	 \langle u(1) \bar{u}({2^\prime}) \rangle &
	 \langle u(1) \bar{u}(3^\prime) \rangle
	 \\
	 \langle u(2) \bar{u}(2^\prime) \rangle &
	 \langle u(2) \bar{u}({3^\prime}) \rangle
	\end{array}
      \right|
  \langle d(4) \bar{d}({5^\prime}) \rangle, 
  \\
 &&
 [\Lambda_{\beta}^{(6)}]
 (\vec{y};c_2^\prime,c_6^\prime,\alpha_6^\prime)
 \nonumber
 \\
 &=&
  \varepsilon(5,6,3)
  \left\{
   (C\gamma_5)(5,6) \delta(\beta,3)
   + (C\gamma_5)(6,3) \delta(\beta,5)
   -2 (C\gamma_5)(3,5) \delta(\beta,6)
  \right\}
  \nonumber
  \\
 && \times 
  \varepsilon(1^\prime,4^\prime,2^\prime)
  (C\gamma_5)(1^\prime,4^\prime) 
  {\langle u(3) \bar{u}({1^\prime}) \rangle}
  \langle d(5) \bar{d}({4^\prime}) \rangle
  \langle s(6) \bar{s}({6^\prime}) \rangle. 
\label{baryonblock_L6_pLLp}
\end{eqnarray}
Note that 
the two outer indices $(\alpha^\prime\beta^\prime)$ in the source 
gather into $[p_{\alpha\alpha^\prime\beta^\prime}^{(6)}]$ because 
the tensorial factors 
$\varepsilon(5^\prime,6^\prime,3^\prime)
{(C\gamma_5)(5^\prime,6^\prime) \delta(\beta^\prime,3^\prime)}$
and 
\\
$\varepsilon(1^\prime,4^\prime,2^\prime)
{(C\gamma_5)(1^\prime,4^\prime)}$
are exchanged between $[p_{\alpha\alpha^\prime\beta^\prime}^{(6)}]$ and 
$[\Lambda_{\beta}^{(6)}]$ 
while ${
\delta(\alpha^\prime,2^\prime)}$ is kept in
$[p_{\alpha\alpha^\prime\beta^\prime}^{(6)}]$. 
The number of explicit summations of indices reduces to 
two colours and one spinor, which makes the 
reduction factor 
$(N_c!N_\alpha)^2\times2^{B-N_\Lambda-N_{\Sigma^0}}/(N_c^2 N_\alpha)= 32$.

\bigskip

Performed these manipulations based on the diagrammatic classification, 
most of the summations can be carried out prior to evaluating the FFT 
so that 
the number of iterations 
significantly reduces;
the numbers of iteration are 
$\{1, 9, 144, 144, 36, 36\}$ 
for the baryon blocks 
$\{([p^{(i)}_\alpha] \times [\Lambda^{(i)}_\beta]); i=1,\cdots,6\}$. 
Therefore only $370$ iterations should be explicitly performed to 
obtain the four-point correlation function of the $p\Lambda$ system 
when we take the operator $\overline{X}_{u}$ 
in $\overline{\Lambda}_{\beta^\prime}$ 
in the source. 
For the sake of completeness, the total number of iterations 
does not change when we take the operator $\overline{X}_{s}$ 
in $\overline{\Lambda}_{\beta^\prime}$ 
in the source 
whereas the numbers of iteration are 
$\{1, 36, 36, 144, 144, 36\}$ when we 
consider the contribution from 
the operator $\overline{X}_{d}$ in $\overline{\Lambda}_{\beta^\prime}$ 
in the source, 
which slightly differ from the former cases 
and the total number of iterations is $397$. 

\section{Extension to 
various $BB$ channels\label{EXTENSION2B8B8CHANNELS}}

The effective block algorithm mentioned above is applicable to 
various $BB$ channels. 
In the recent few years, the 2+1 flavour lattice QCD calculations have 
been widely performed. 
This is an opportune moment to 
go beyond 
the $BB$ potentials 
at the flavour $SU(3)$ point~\cite{Inoue:2011ai} 
since 
exploring breakdown of the flavour symmetry is 
not only an intriguing subject but also 
a 
major 
concern 
of the phenomenological $YN$ and $YY$ interaction models. 
Therefore, it is beneficial to take account of 
a large number of $BB$ channels. 
For example, 
we 
consider 
the following $52$ 
four-point correlation functions 
in order to study the complete set of $BB$ interactions 
in the isospin symmetric limit. 
(For the moment, we assume that 
the electromagnetic interaction is not taken into account 
in the present lattice calculation.)

{\footnotesize
\begin{eqnarray}
&&
\!\!\!\!
\!\!\!\!\!\!\!\!
\langle pn\overline{pn}\rangle,~ 
\label{GeneralBB_NN}
\\
&&
\!\!\!\!\!\!\!\!\!\!\!\!\!\!\!\!
\begin{array}{lll}
\langle p\Lambda\overline{p\Lambda}\rangle, &  \langle p\Lambda\overline{\Sigma^{+}n}\rangle, &  \langle p\Lambda\overline{\Sigma^{0}p}\rangle, 
\\
\langle \Sigma^{+}n\overline{p\Lambda}\rangle, &  \langle \Sigma^{+}n\overline{\Sigma^{+}n}\rangle, &  \langle \Sigma^{+}n\overline{\Sigma^{0}p}\rangle, 
\\
\langle \Sigma^{0}p\overline{p\Lambda}\rangle, &  \langle \Sigma^{0}p\overline{\Sigma^{+}n}\rangle, &  \langle \Sigma^{0}p\overline{\Sigma^{0}p}\rangle, 
\end{array}
\label{GeneralBB_NL}
\\
&&
\!\!\!\!\!\!\!\!\!\!\!\!\!\!\!\!
\begin{array}{llllll}
\langle \Lambda\Lambda\overline{\Lambda\Lambda}\rangle, &\!\!\!\! \langle \Lambda\Lambda\overline{p\Xi^{-}}\rangle, &\!\!\!\! \langle \Lambda\Lambda\overline{n\Xi^{0}}\rangle, &\!\!\!\! \langle \Lambda\Lambda\overline{\Sigma^{+}\Sigma^{-}}\rangle, &\!\!\!\! \langle \Lambda\Lambda\overline{\Sigma^{0}\Sigma^{0}}\rangle, 
\\
\langle p\Xi^{-}\overline{\Lambda\Lambda}\rangle, &\!\!\!\! \langle p\Xi^{-}\overline{p\Xi^{-}}\rangle, &\!\!\!\! \langle p\Xi^{-}\overline{n\Xi^{0}}\rangle, &\!\!\!\! \langle p\Xi^{-}\overline{\Sigma^{+}\Sigma^{-}}\rangle, &\!\!\!\! \langle p\Xi^{-}\overline{\Sigma^{0}\Sigma^{0}}\rangle, &\!\!\!\! \langle p\Xi^{-}\overline{\Sigma^{0}\Lambda}\rangle,~
\\
\langle n\Xi^{0}\overline{\Lambda\Lambda}\rangle, &\!\!\!\! \langle n\Xi^{0}\overline{p\Xi^{-}}\rangle, &\!\!\!\! \langle n\Xi^{0}\overline{n\Xi^{0}}\rangle, &\!\!\!\! \langle n\Xi^{0}\overline{\Sigma^{+}\Sigma^{-}}\rangle, &\!\!\!\! \langle n\Xi^{0}\overline{\Sigma^{0}\Sigma^{0}}\rangle, &\!\!\!\! \langle n\Xi^{0}\overline{\Sigma^{0}\Lambda}\rangle,~
\\
\langle \Sigma^{+}\Sigma^{-}\overline{\Lambda\Lambda}\rangle, &\!\!\!\! \langle \Sigma^{+}\Sigma^{-}\overline{p\Xi^{-}}\rangle, &\!\!\!\! \langle \Sigma^{+}\Sigma^{-}\overline{n\Xi^{0}}\rangle, &\!\!\!\! \langle \Sigma^{+}\Sigma^{-}\overline{\Sigma^{+}\Sigma^{-}}\rangle, &\!\!\!\! \langle \Sigma^{+}\Sigma^{-}\overline{\Sigma^{0}\Sigma^{0}}\rangle, &\!\!\!\! \langle \Sigma^{+}\Sigma^{-}\overline{\Sigma^{0}\Lambda}\rangle,~
\\
\langle \Sigma^{0}\Sigma^{0}\overline{\Lambda\Lambda}\rangle, &\!\!\!\! \langle \Sigma^{0}\Sigma^{0}\overline{p\Xi^{-}}\rangle, &\!\!\!\! \langle \Sigma^{0}\Sigma^{0}\overline{n\Xi^{0}}\rangle, &\!\!\!\! \langle \Sigma^{0}\Sigma^{0}\overline{\Sigma^{+}\Sigma^{-}}\rangle, &\!\!\!\! \langle \Sigma^{0}\Sigma^{0}\overline{\Sigma^{0}\Sigma^{0}}\rangle,~
\\
&\!\!\!\! \langle \Sigma^{0}\Lambda\overline{p\Xi^{-}}\rangle, &\!\!\!\! \langle \Sigma^{0}\Lambda\overline{n\Xi^{0}}\rangle, &\!\!\!\! \langle \Sigma^{0}\Lambda\overline{\Sigma^{+}\Sigma^{-}}\rangle, & &\!\!\!\! \langle \Sigma^{0}\Lambda\overline{\Sigma^{0}\Lambda}\rangle, 
\end{array}
\label{GeneralBB_LL}
\\
&&
\!\!\!\!\!\!\!\!\!\!\!\!\!\!\!\!
\begin{array}{lll}
\langle \Xi^{-}\Lambda\overline{\Xi^{-}\Lambda}\rangle, & \langle \Xi^{-}\Lambda\overline{\Sigma^{-}\Xi^{0}}\rangle, & \langle \Xi^{-}\Lambda\overline{\Sigma^{0}\Xi^{-}}\rangle,~
\\
\langle \Sigma^{-}\Xi^{0}\overline{\Xi^{-}\Lambda}\rangle, & \langle \Sigma^{-}\Xi^{0}\overline{\Sigma^{-}\Xi^{0}}\rangle, & \langle \Sigma^{-}\Xi^{0}\overline{\Sigma^{0}\Xi^{-}}\rangle,~
\\
\langle \Sigma^{0}\Xi^{-}\overline{\Xi^{-}\Lambda}\rangle, & \langle \Sigma^{0}\Xi^{-}\overline{\Sigma^{-}\Xi^{0}}\rangle, & \langle \Sigma^{0}\Xi^{-}\overline{\Sigma^{0}\Xi^{-}}\rangle,~
\end{array}
\label{GeneralBB_XL}
\\
&&
\!\!\!\!
\!\!\!\!\!\!\!\!
\langle \Xi^{-}\Xi^{0}\overline{\Xi^{-}\Xi^{0}}\rangle,~
\label{GeneralBB_XX}
\end{eqnarray}
}
We omit four off-diagonal channels, 
$\langle \Lambda\Lambda\overline{\Sigma^{0}\Lambda}\rangle$, 
$\langle \Sigma^{0}\Sigma^{0}\overline{\Sigma^{0}\Lambda}\rangle$, 
$\langle \Sigma^{0}\Lambda\overline{\Lambda\Lambda}\rangle$ 
and 
$\langle \Sigma^{0}\Lambda\overline{\Sigma^{0}\Sigma^{0}}\rangle$, 
from the above list 
because they are expected to be identically zero 
in the isospin symmetric limit~\footnote{
In this paper, we focus on the $2+1$ flavour lattice QCD calculation 
for the study of the octet-baryon-octet-baryon interactions in the 
isospin symmetric limit. 
An extension to the other charge states than the channels given in 
Eqs.~(\ref{GeneralBB_NN})$-$(\ref{GeneralBB_XX}) is straightforward. 
Moreover, 
even though the system 
comprises 
decuplet baryons such as $\Omega^{-}$'s, 
we can take Eq.~(\ref{FieldOperator_B}) and the 
gamma matrices 
$\Gamma_{1}=C\gamma_{\ell}$ and $\Gamma_{2}=1$ 
with spatial vector index $\ell$. 
}. 
For an extension of the calculation of the four-point correlator 
to various $BB$ channels, 
we have implemented a C++ program to perform the Wick contraction 
together with the FFT 
in terms of the diagrammatic classification, 
the procedures of which are automatically performed 
once the interpolating fields in 
the source and sink 
(i.e., the quantum numbers of the system) are given. 
We also independently implemented another C++ program which performs the 
Wick contractions to calculate the above 52 channels of 
four-point correlation function without employing the FFT. 
We have confirmed that the numerical results obtained by 
the present effective block algorithm agree with the numerical 
results calculated by the latter C++ program. 
See also 
Sec.~\ref{BENCHMARK} for thoroughgoing check between 
this algorithm and the unified contraction algorithm.

Table~\ref{NumberOfIterationsS01} lists 
the number of diagrams, 
the number of iterations together with the parity of the permutation, 
and the number of total iterations for the four-point correlation 
functions of various $BB$ channels with the strangeness 
$S=0$ and $-1$. 
For the $NN$ system, 
the number of total iterations for the 
channel $\langle pn \overline{pn}\rangle$ 
is just 586, 
which is quite small 
in comparison with the $N_{\rm contr}=2358$ 
in Table A.3 in Ref.~\cite{Doi:2012xd}. 
As is discussed in the previous section, 
for the channels with $S=-1$, 
the numbers of iteration 
lessened by the effective block algorithm 
depend on the form of the diquark combination 
in the baryon field operators in the source. 
It is therefore convenient to separate between the contributions 
from the fields 
$\overline{X}_{u}$, $\overline{X}_{d}$, and $\overline{X}_{s}$ 
if the correlator comprises the field(s) 
$\overline{\Lambda}$ and/or 
$\overline{\Sigma^{0}}$ in the source. 
In the Table~\ref{NumberOfIterationsS01}, 
we explicitly indicate the form of diquark combination such as 
$\langle p\Lambda\overline{p\Lambda_{X_{u,s}}}\rangle$ and 
$\langle p\Lambda\overline{p\Lambda_{X_d}}\rangle$ 
to distinguish the diquark combinations when the correlator 
includes $\overline{\Lambda}$ and/or $\overline{\Sigma^{0}}$. 
Among the numbers of the total iterations 
for various channels with the strangeness $S=-1$, 
the largest number is 405 which is found at the channels of 
$\langle p\Lambda \overline{\Sigma^{+}n}\rangle$ and 
$\langle \Sigma^{0}p \overline{\Sigma^{+}n}\rangle$; 
it is noticeably smaller than the smallest value $N_{\rm contr}=1350$ 
(except 0) 
among the Tables A.1, A.3 and A.5 
in Ref.~\cite{Doi:2012xd}.

Because there are a lot of channels for the strangeness $S=-2$, 
we divide the list into two parts. 
Table~\ref{NumberOfIterationsS2_0} 
(Table~\ref{NumberOfIterationsS2_1}) 
shows the first (second) part of the list of the numbers of iteration 
for the channels with the strangeness $S=-2$. 
The five four-point correlation functions, 
$\langle \Lambda\Lambda\overline{\Lambda_{X_q}\Lambda_{X_{q^\prime}}}\rangle$, 
$\langle \Lambda\Lambda\overline{\Sigma^{0}_{X_q}\Sigma^{0}_{X_{q^\prime}}}\rangle$, 
$\langle \Sigma^{0}\Sigma^{0}\overline{\Lambda_{X_q}\Lambda_{X_{q^\prime}}}\rangle$, 
$\langle \Sigma^{0}\Sigma^{0}\overline{\Sigma^{0}_{X_q}\Sigma^{0}_{X_{q^\prime}}}\rangle$, 
and 
$\langle \Sigma^{0}\Lambda\overline{\Sigma^{0}_{X_q}\Lambda_{X_{q^\prime}}}\rangle$ 
(for $q=q^\prime$), 
are the relatively computationally demanding channels 
in the Tables; 
the total numbers of iterations are all 596 for these channels and 
they are remarkably smaller than the any $N_{\rm contr}$ values 
(except 0) 
among the Tables A.1, A.3 and A.5 
in Ref.~\cite{Doi:2012xd}.

Table~\ref{NumberOfIterationsS34} shows the numbers of iterations 
to calculate the four-point correlation functions of the 
strangeness $S=-3$ and $-4$ systems. 
There are similarities in the list of numbers between $S=-3$ and 
$S=-1$ since the isospin quantum number of $\Xi$ is same as the 
isospin of $N$. 
Therefore the efficiency for the calculation of correlators 
of $S=-3$ systems is similar to that of $S=-1$ systems. 
On the other hand, the numbers of iterations to calculate the 
four-point correlation function of the $S=-4$ system differ 
from the numbers of iterations to calculate the correlator of 
the $S=0$ system. 
The total number of iterations is $370$ for 
$\langle \Xi^{-}\Xi^{0}\overline{\Xi^{-}\Xi^{0}}\rangle$
whereas 
the total number of iterations is $586$ for $\langle pn\overline{pn}\rangle$.

\begin{table}[t]
 \begin{minipage}{\textwidth}
  \begin{center}
   \caption{
The number of diagrams, 
the number of iterations together with the parity of the permutation 
for each diagram, 
and the number of total iterations for the four-point correlation 
functions of various $BB$ channels with the strangeness 
$S=0$ and $-1$. 
See text for details. 
\label{NumberOfIterationsS01}}
  \footnotesize
   \begin{tabular}{cccc}
    \hline
channel & {\scriptsize{\shortstack{ \# of \\diagrams}}} & \{(\# of iterations)$^{\rm sign}$\} & {\scriptsize{\shortstack{\# of total \\iterations}}} \\
    \hline
    %
%
 $\langle pn\overline{pn}\rangle$ & 9 & $\{   1^{+},  36^{-}, 144^{-},  36^{+},  36^{+}, 144^{-}, 144^{+},   9^{-},  36^{+} \}$ & 586 \\
 $\langle p\Lambda\overline{p\Lambda_{X_{u,s}}}\rangle$ & 6 & $\{   1^{+},   9^{-}, 144^{-}, 144^{+},  36^{+},  36^{-} \}$ & 370 \\
 $\langle p\Lambda\overline{p\Lambda_{X_d}}\rangle$ & 6 & $\{   1^{+},  36^{-},  36^{-}, 144^{+}, 144^{+},  36^{-} \}$ & 397 \\
 $\langle p\Lambda\overline{\Sigma^{+}n}\rangle$ & 6 & $\{ 144^{-},  36^{+},  36^{+}, 144^{-},   9^{-},  36^{+} \}$ & 405 \\
 $\langle p\Lambda\overline{\Sigma^{0}_{X_u}p}\rangle$ & 6 & $\{ 144^{+},  36^{-},   9^{-},  36^{+}, 144^{+},   1^{-} \}$ & 370 \\
 $\langle p\Lambda\overline{\Sigma^{0}_{X_d}p}\rangle$ & 6 & $\{ 144^{-},  36^{+},  36^{+}, 144^{-},  36^{-},   1^{+} \}$ & 397 \\
 $\langle \Sigma^{+}n\overline{p\Lambda_{X_u}}\rangle$ & 3 & $\{ 144^{-}, 144^{+},  36^{-} \}$ & 324 \\
 $\langle \Sigma^{+}n\overline{p\Lambda_{X_d}}\rangle$ & 3 & $\{ 144^{-},  36^{+},   9^{-} \}$ & 189 \\
 $\langle \Sigma^{+}n\overline{p\Lambda_{X_s}}\rangle$ & 3 & $\{  36^{-}, 144^{+},  36^{-} \}$ & 216 \\
 $\langle \Sigma^{+}n\overline{\Sigma^{+}n}\rangle$ & 3 & $\{   1^{+},  36^{-}, 144^{+} \}$ & 181 \\
 $\langle \Sigma^{+}n\overline{\Sigma^{0}_{X_u}p}\rangle$ & 3 & $\{ 144^{-},  36^{+}, 144^{-} \}$ & 324 \\
 $\langle \Sigma^{+}n\overline{\Sigma^{0}_{X_d}p}\rangle$ & 3 & $\{  36^{+},   9^{-}, 144^{+} \}$ & 189 \\
 $\langle \Sigma^{0}p\overline{p\Lambda_{X_{u,s}}}\rangle$ & 6 & $\{  36^{+}, 144^{-}, 144^{+},  36^{-},   9^{+},   1^{-} \}$ & 370 \\
 $\langle \Sigma^{0}p\overline{p\Lambda_{X_d}}\rangle$ & 6 & $\{  36^{+}, 144^{-},  36^{+}, 144^{-},  36^{+},   1^{-} \}$ & 397 \\
 $\langle \Sigma^{0}p\overline{\Sigma^{+}n}\rangle$ & 6 & $\{  36^{-}, 144^{+},  36^{-},   9^{+},  36^{-}, 144^{+} \}$ & 405 \\
 $\langle \Sigma^{0}p\overline{\Sigma^{0}_{X_u}p}\rangle$ & 6 & $\{   1^{+},  36^{-},   9^{+}, 144^{-},  36^{+}, 144^{-} \}$ & 370 \\
 $\langle \Sigma^{0}p\overline{\Sigma^{0}_{X_d}p}\rangle$ & 6 & $\{   1^{-}, 144^{+},  36^{-},  36^{+},  36^{-}, 144^{+} \}$ & 397 \\
%
    %
    \hline
   \end{tabular}
  \end{center}
 \end{minipage}
\end{table}

\begin{table}[t]
 \begin{minipage}{\textwidth}
  \begin{center}
   \caption{
Same as Table~\ref{NumberOfIterationsS01} but for the first part of 
channels with the strangeness $S=-2$. 
\label{NumberOfIterationsS2_0}}
  \footnotesize
   \begin{tabular}{lccc}
    \hline
channel & {\scriptsize{\shortstack{ \# of \\diagrams}}} & \{(\# of iterations)$^{\rm sign}$\} & {\scriptsize{\shortstack{ \# of total \\iterations}}} \\
    \hline
    %
%
 $\langle \Lambda\Lambda\overline{\Lambda_{X_q}\Lambda_{X_{q^\prime}}}\rangle$ $(q=q^\prime)$ & 8 & $\{   1^{+},   9^{-}, 144^{-}, 144^{+}, 144^{-}, 144^{+},   9^{+},   1^{-} \}$ & 596 \\
 $\langle \Lambda\Lambda\overline{\Lambda_{X_q}\Lambda_{X_{q^\prime}}}\rangle$ $(q\ne q^\prime)$ & 8 & $\{   1^{+},  36^{-}, 144^{-},  36^{+},  36^{-}, 144^{+},  36^{+},   1^{-} \}$ & 434 \\
 $\langle \Lambda\Lambda\overline{p\Xi^{-}}\rangle$ & 8 & $\{  36^{+}, 144^{-},   9^{-},  36^{+},  36^{-},   9^{+}, 144^{+},  36^{-} \}$ & 450 \\
 $\langle \Lambda\Lambda\overline{n\Xi^{0}}\rangle$ & 8 & $\{  36^{+},  36^{-},   9^{-}, 144^{+}, 144^{-},   9^{+},  36^{+},  36^{-} \}$ & 450 \\
 $\langle \Lambda\Lambda\overline{\Sigma^{+}\Sigma^{-}}\rangle$ & 8 & $\{  36^{-}, 144^{+},  36^{+},   9^{-},   9^{+},  36^{-}, 144^{-},  36^{+} \}$ & 450 \\
 $\langle \Lambda\Lambda\overline{\Sigma^{0}_{X_q}\Sigma^{0}_{X_{q^\prime}}}\rangle$ $(q=q^\prime)$ & 8 & $\{   1^{+},   9^{-}, 144^{-}, 144^{+}, 144^{-}, 144^{+},   9^{+},   1^{-} \}$ & 596 \\
 $\langle \Lambda\Lambda\overline{\Sigma^{0}_{X_q}\Sigma^{0}_{X_{q^\prime}}}\rangle$ $(q\ne q^\prime)$ & 8 & $\{   1^{-},  36^{+}, 144^{+},  36^{-},  36^{+}, 144^{-},  36^{-},   1^{+} \}$ & 434 \\
 $\langle p\Xi^{-}\overline{\Lambda_{X_q}\Lambda_{X_q}}\rangle$ $(q=u,s)$ & 2 & $\{  36^{+},  36^{-} \}$ & 72 \\
 $\langle p\Xi^{-}\overline{\Lambda_{X_q}\Lambda_{X_{q^\prime}}}\rangle$ \\ \quad $^{((q,q^\prime)=(d,u),(u,d),(s,d),(d,s))}$ & 2 & $\{  36^{+}, 144^{-} \}$ & 180 \\
 $\langle p\Xi^{-}\overline{\Lambda_{X_q}\Lambda_{X_{q^\prime}}}\rangle$ \\ \quad $^{((q,q^\prime)=(s,u),(u,s))}$ & 2 & $\{   9^{+}, 144^{-} \}$ & 153 \\
 $\langle p\Xi^{-}\overline{\Lambda_{X_d}\Lambda_{X_d}}\rangle$ & 2 & $\{ 144^{+}, 144^{-} \}$ & 288 \\
 $\langle p\Xi^{-}\overline{p\Xi^{-}}\rangle$ & 2 & $\{   1^{+}, 144^{-} \}$ & 145 \\
 $\langle p\Xi^{-}\overline{n\Xi^{0}}\rangle$ & 2 & $\{  36^{+}, 144^{-} \}$ & 180 \\
 $\langle p\Xi^{-}\overline{\Sigma^{+}\Sigma^{-}}\rangle$ & 2 & $\{ 144^{-},  36^{+} \}$ & 180 \\
 $\langle p\Xi^{-}\overline{\Sigma^{0}_{X_u}\Sigma^{0}_{X_u}}\rangle$ & 2 & $\{  36^{+},  36^{-} \}$ & 72 \\
 $\langle p\Xi^{-}\overline{\Sigma^{0}_{X_q}\Sigma^{0}_{X_{q^\prime}}}\rangle$ $(q\ne q^\prime)$ & 2 & $\{  36^{-}, 144^{+} \}$ & 180 \\
 $\langle p\Xi^{-}\overline{\Sigma^{0}_{X_d}\Sigma^{0}_{X_d}}\rangle$ & 2 & $\{ 144^{+}, 144^{-} \}$ & 288 \\
 $\langle p\Xi^{-}\overline{\Sigma^{0}_{X_u}\Lambda_{X_u}}\rangle$ & 2 & $\{  36^{+},  36^{-} \}$ & 72 \\
 $\langle p\Xi^{-}\overline{\Sigma^{0}_{X_q}\Lambda_{X_{q^\prime}}}\rangle$ \\ \quad $^{((q,q^\prime)=(d,u),(u,d),(d,s))}$ & 2 & $\{  36^{-}, 144^{+} \}$ & 180 \\
 $\langle p\Xi^{-}\overline{\Sigma^{0}_{X_d}\Lambda_{X_d}}\rangle$ & 2 & $\{ 144^{-}, 144^{+} \}$ & 288 \\
 $\langle p\Xi^{-}\overline{\Sigma^{0}_{X_u}\Lambda_{X_s}}\rangle$ & 2 & $\{ 144^{+},   9^{-} \}$ & 153 \\
 $\langle n\Xi^{0}\overline{\Lambda_{X_u}\Lambda_{X_u}}\rangle$ & 2 & $\{ 144^{+}, 144^{-} \}$ & 288 \\
 $\langle n\Xi^{0}\overline{\Lambda_{X_q}\Lambda_{X_{q^\prime}}}\rangle$ \\ \quad $^{((q,q^\prime)=(d,u),(u,d),(s,u),(u,s))}$ & 2 & $\{ 144^{+},  36^{-} \}$ & 180 \\
 $\langle n\Xi^{0}\overline{\Lambda_{X_q}\Lambda_{X_q}}\rangle$ $(q=d,s)$ & 2 & $\{  36^{+},  36^{-} \}$ & 72 \\
 $\langle n\Xi^{0}\overline{\Lambda_{X_q}\Lambda_{X_{q^\prime}}}\rangle$ \\ \quad $^{((q,q^\prime)=(s,d),(d,s))}$ & 2 & $\{   9^{+}, 144^{-} \}$ & 153 \\
 $\langle n\Xi^{0}\overline{p\Xi^{-}}\rangle$ & 2 & $\{  36^{+}, 144^{-} \}$ & 180 \\
 $\langle n\Xi^{0}\overline{n\Xi^{0}}\rangle$ & 2 & $\{   1^{+}, 144^{-} \}$ & 145 \\
 $\langle n\Xi^{0}\overline{\Sigma^{+}\Sigma^{-}}\rangle$ & 2 & $\{ 144^{-},  36^{+} \}$ & 180 \\
 $\langle n\Xi^{0}\overline{\Sigma^{0}_{X_u}\Sigma^{0}_{X_u}}\rangle$ & 2 & $\{ 144^{+}, 144^{-} \}$ & 288 \\
 $\langle n\Xi^{0}\overline{\Sigma^{0}_{X_q}\Sigma^{0}_{X_{q^\prime}}}\rangle$ ($q\ne q^\prime$) & 2 & $\{ 144^{-},  36^{+} \}$ & 180 \\
 $\langle n\Xi^{0}\overline{\Sigma^{0}_{X_d}\Sigma^{0}_{X_d}}\rangle$ & 2 & $\{  36^{+},  36^{-} \}$ & 72 \\
 $\langle n\Xi^{0}\overline{\Sigma^{0}_{X_u}\Lambda_{X_u}}\rangle$ & 2 & $\{ 144^{+}, 144^{-} \}$ & 288 \\
 $\langle n\Xi^{0}\overline{\Sigma^{0}_{X_q}\Lambda_{X_{q^\prime}}}\rangle$ \\ \quad $^{((q,q^\prime)=(d,u),(u,d),(u,s))}$ & 2 & $\{ 144^{-},  36^{+} \}$ & 180 \\
 $\langle n\Xi^{0}\overline{\Sigma^{0}_{X_d}\Lambda_{X_d}}\rangle$ & 2 & $\{  36^{-},  36^{+} \}$ & 72 \\
 $\langle n\Xi^{0}\overline{\Sigma^{0}_{X_d}\Lambda_{X_s}}\rangle$ & 2 & $\{ 144^{-},   9^{+} \}$ & 153 \\
%
    %
    \hline
   \end{tabular}
  \end{center}
 \end{minipage}
\end{table}

\begin{table}[t]
 \begin{minipage}{\textwidth}
  \begin{center}
   \caption{
Same as Table~\ref{NumberOfIterationsS01} but for the second part of 
channels with the strangeness $S=-2$. 
\label{NumberOfIterationsS2_1}}
  \footnotesize
   \begin{tabular}{lccc}
    \hline
channel & {\scriptsize{\shortstack{ \# of \\diagrams}}} & \{(\# of iterations)$^{\rm sign}$\} & {\scriptsize{\shortstack{ \# of total \\iterations}}} \\
    \hline
    %
%
 $\langle \Sigma^{+}\Sigma^{-}\overline{\Lambda_{X_q}\Lambda_{X_q}}\rangle$ $(q=u,d)$ & 2 & $\{  36^{-},  36^{+} \}$ & 72 \\
 $\langle \Sigma^{+}\Sigma^{-}\overline{\Lambda_{X_q}\Lambda_{X_{q^\prime}}}\rangle$ \\ \quad $^{((q,q^\prime)=(d,u),(u,d))}$ & 2 & $\{   9^{-}, 144^{+} \}$ & 153 \\
 $\langle \Sigma^{+}\Sigma^{-}\overline{\Lambda_{X_q}\Lambda_{X_{q^\prime}}}\rangle$ \\ \quad $^{((q,q^\prime)=(s,u),(s,d),(u,s),(d,s))}$ & 2 & $\{  36^{-}, 144^{+} \}$ & 180 \\
 $\langle \Sigma^{+}\Sigma^{-}\overline{\Lambda_{X_s}\Lambda_{X_s}}\rangle$ & 2 & $\{ 144^{-}, 144^{+} \}$ & 288 \\
 $\langle \Sigma^{+}\Sigma^{-}\overline{p\Xi^{-}}\rangle$ & 2 & $\{ 144^{-},  36^{+} \}$ & 180 \\
 $\langle \Sigma^{+}\Sigma^{-}\overline{n\Xi^{0}}\rangle$ & 2 & $\{  36^{-}, 144^{+} \}$ & 180 \\
 $\langle \Sigma^{+}\Sigma^{-}\overline{\Sigma^{+}\Sigma^{-}}\rangle$ & 2 & $\{   1^{+}, 144^{-} \}$ & 145 \\
 $\langle \Sigma^{+}\Sigma^{-}\overline{\Sigma^{0}_{X_q}\Sigma^{0}_{X_q}}\rangle$ $(q=u,d)$ & 2 & $\{  36^{-},  36^{+} \}$ & 72 \\
 $\langle \Sigma^{+}\Sigma^{-}\overline{\Sigma^{0}_{X_q}\Sigma^{0}_{X_{q^\prime}}}\rangle$ $(q\ne q^\prime)$ & 2 & $\{   9^{+}, 144^{-} \}$ & 153 \\
 $\langle \Sigma^{+}\Sigma^{-}\overline{\Sigma^{0}_{X_q}\Lambda_{X_q}}\rangle$ $(q=u,d)$ & 2 & $\{  36^{-},  36^{+} \}$ & 72 \\
 $\langle \Sigma^{+}\Sigma^{-}\overline{\Sigma^{0}_{X_q}\Lambda_{X_{q^\prime}}}\rangle$ \\ \quad $^{((q,q^\prime)=(d,u),(u,d))}$ & 2 & $\{   9^{+}, 144^{-} \}$ & 153 \\
 $\langle \Sigma^{+}\Sigma^{-}\overline{\Sigma^{0}_{X_q}\Lambda_{X_s}}\rangle$ $(q=u,d)$ & 2 & $\{ 144^{-},  36^{+} \}$ & 180 \\
 $\langle \Sigma^{0}\Sigma^{0}\overline{\Lambda_{X_q}\Lambda_{X_{q^\prime}}}\rangle$ $(q=q^\prime)$ & 8 & $\{   1^{+},   9^{-}, 144^{-}, 144^{+}, 144^{-}, 144^{+},   9^{+},   1^{-} \}$ & 596 \\
 $\langle \Sigma^{0}\Sigma^{0}\overline{\Lambda_{X_q}\Lambda_{X_{q^\prime}}}\rangle$ $(q\ne q^\prime)$ & 8 & $\{   1^{+},  36^{-}, 144^{-},  36^{+},  36^{-}, 144^{+},  36^{+},   1^{-} \}$ & 434 \\
 $\langle \Sigma^{0}\Sigma^{0}\overline{p\Xi^{-}}\rangle$ & 8 & $\{  36^{+}, 144^{-},   9^{-},  36^{+},  36^{-},   9^{+}, 144^{+},  36^{-} \}$ & 450 \\
 $\langle \Sigma^{0}\Sigma^{0}\overline{n\Xi^{0}}\rangle$ & 8 & $\{  36^{+},  36^{-},   9^{-}, 144^{+}, 144^{-},   9^{+},  36^{+},  36^{-} \}$ & 450 \\
 $\langle \Sigma^{0}\Sigma^{0}\overline{\Sigma^{+}\Sigma^{-}}\rangle$ & 8 & $\{  36^{-}, 144^{+},  36^{+},   9^{-},   9^{+},  36^{-}, 144^{-},  36^{+} \}$ & 450 \\
 $\langle \Sigma^{0}\Sigma^{0}\overline{\Sigma^{0}_{X_q}\Sigma^{0}_{X_{q^\prime}}}\rangle$ $(q=q^\prime)$ & 8 & $\{   1^{+},   9^{-}, 144^{-}, 144^{+}, 144^{-}, 144^{+},   9^{+},   1^{-} \}$ & 596 \\
 $\langle \Sigma^{0}\Sigma^{0}\overline{\Sigma^{0}_{X_q}\Sigma^{0}_{X_{q^\prime}}}\rangle$ $(q\ne q^\prime)$ & 8 & $\{   1^{-},  36^{+}, 144^{+},  36^{-},  36^{+}, 144^{-},  36^{-},   1^{+} \}$ & 434 \\
 $\langle \Sigma^{0}\Lambda\overline{p\Xi^{-}}\rangle$ & 8 & $\{  36^{+}, 144^{-},   9^{-},  36^{+},  36^{-},   9^{+}, 144^{+},  36^{-} \}$ & 450 \\
 $\langle \Sigma^{0}\Lambda\overline{n\Xi^{0}}\rangle$ & 8 & $\{  36^{+},  36^{-},   9^{-}, 144^{+}, 144^{-},   9^{+},  36^{+},  36^{-} \}$ & 450 \\
 $\langle \Sigma^{0}\Lambda\overline{\Sigma^{+}\Sigma^{-}}\rangle$ & 8 & $\{  36^{-}, 144^{+},  36^{+},   9^{-},   9^{+},  36^{-}, 144^{-},  36^{+} \}$ & 450 \\
 $\langle \Sigma^{0}\Lambda\overline{\Sigma^{0}_{X_q}\Lambda_{X_{q^\prime}}}\rangle$ $(q=q^\prime)$ & 8 & $\{   1^{+},   9^{-}, 144^{-}, 144^{+}, 144^{-}, 144^{+},   9^{+},   1^{-} \}$ & 596 \\
 $\langle \Sigma^{0}\Lambda\overline{\Sigma^{0}_{X_q}\Lambda_{X_{q^\prime}}}\rangle$ $(q\ne q^\prime)$ & 8 & $\{   1^{-},  36^{+}, 144^{+},  36^{-},  36^{+}, 144^{-},  36^{-},   1^{+} \}$ & 434 \\
%
    %
    \hline
   \end{tabular}
  \end{center}
 \end{minipage}
\end{table}

\begin{table}[t]
 \begin{minipage}{\textwidth}
  \begin{center}
   \caption{
Same as Table~\ref{NumberOfIterationsS01} but for the channels 
with the strangeness $S=-3$ and $-4$. 
\label{NumberOfIterationsS34}}
  \footnotesize
   \begin{tabular}{cccc}
    \hline
channel & {\scriptsize{\shortstack{ \# of \\diagrams}}} & \{(\# of iterations)$^{\rm sign}$\} & {\scriptsize{\shortstack{ \# of total \\iterations}}} \\
    \hline
    %
%
 $\langle \Xi^{-}\Lambda\overline{\Xi^{-}\Lambda_{X_{u,s}}}\rangle$ & 6 & $\{   1^{+},  36^{-}, 144^{+}, 144^{-},  36^{+},   9^{-} \}$ & 370 \\
 $\langle \Xi^{-}\Lambda\overline{\Xi^{-}\Lambda_{X_d}}\rangle$ & 6 & $\{   1^{+},  36^{-}, 144^{+},  36^{-}, 144^{+},  36^{-} \}$ & 397 \\
 $\langle \Xi^{-}\Lambda\overline{\Sigma^{-}\Xi^{0}}\rangle$ & 6 & $\{  36^{-},   9^{+}, 144^{-}, 144^{+},  36^{-},  36^{+} \}$ & 405 \\
 $\langle \Xi^{-}\Lambda\overline{\Sigma^{0}_{X_u}\Xi^{-}}\rangle$ & 6 & $\{  36^{+},   9^{-}, 144^{+},  36^{-}, 144^{+},   1^{-} \}$ & 370 \\
 $\langle \Xi^{-}\Lambda\overline{\Sigma^{0}_{X_d}\Xi^{-}}\rangle$ & 6 & $\{ 144^{-},  36^{+},  36^{-},  36^{+}, 144^{-},   1^{+} \}$ & 397 \\
 $\langle \Sigma^{-}\Xi^{0}\overline{\Xi^{-}\Lambda_{X_u}}\rangle$ & 3 & $\{  36^{-}, 144^{+},  36^{-} \}$ & 216 \\
 $\langle \Sigma^{-}\Xi^{0}\overline{\Xi^{-}\Lambda_{X_d}}\rangle$ & 3 & $\{   9^{-},  36^{+}, 144^{-} \}$ & 189 \\
 $\langle \Sigma^{-}\Xi^{0}\overline{\Xi^{-}\Lambda_{X_s}}\rangle$ & 3 & $\{  36^{-}, 144^{+}, 144^{-} \}$ & 324 \\
 $\langle \Sigma^{-}\Xi^{0}\overline{\Sigma^{-}\Xi^{0}}\rangle$ & 3 & $\{   1^{+}, 144^{-},  36^{+} \}$ & 181 \\
 $\langle \Sigma^{-}\Xi^{0}\overline{\Sigma^{0}_{X_u}\Xi^{-}}\rangle$ & 3 & $\{  36^{-},  36^{+}, 144^{-} \}$ & 216 \\
 $\langle \Sigma^{-}\Xi^{0}\overline{\Sigma^{0}_{X_d}\Xi^{-}}\rangle$ & 3 & $\{ 144^{+},   9^{-},  36^{+} \}$ & 189 \\
 $\langle \Sigma^{0}\Xi^{-}\overline{\Xi^{-}\Lambda_{X_{u,s}}}\rangle$ & 6 & $\{   9^{+},  36^{-}, 144^{+}, 144^{-},  36^{+},   1^{-} \}$ & 370 \\
 $\langle \Sigma^{0}\Xi^{-}\overline{\Xi^{-}\Lambda_{X_d}}\rangle$ & 6 & $\{  36^{+}, 144^{-},  36^{+}, 144^{-},  36^{+},   1^{-} \}$ & 397 \\
 $\langle \Sigma^{0}\Xi^{-}\overline{\Sigma^{-}\Xi^{0}}\rangle$ & 6 & $\{  36^{-},  36^{+}, 144^{-}, 144^{+},   9^{-},  36^{+} \}$ & 405 \\
 $\langle \Sigma^{0}\Xi^{-}\overline{\Sigma^{0}_{X_u}\Xi^{-}}\rangle$ & 6 & $\{   1^{+}, 144^{-},  36^{+}, 144^{-},   9^{+},  36^{-} \}$ & 370 \\
 $\langle \Sigma^{0}\Xi^{-}\overline{\Sigma^{0}_{X_d}\Xi^{-}}\rangle$ & 6 & $\{   1^{-}, 144^{+},  36^{-},  36^{+},  36^{-}, 144^{+} \}$ & 397 \\
 $\langle \Xi^{-}\Xi^{0}\overline{\Xi^{-}\Xi^{0}}\rangle$ & 6 & $\{   1^{+},  36^{-},   9^{+}, 144^{+},  36^{-}, 144^{+} \}$ & 370 \\
%
    %
    \hline
   \end{tabular}
  \end{center}
 \end{minipage}
\end{table}

\section{
Hybrid parallel computation of the four-point 
correlators 
\label{IMPLEMENTATION}}

%
The message passing interface (MPI) is a message-passing standard 
designed for distributed memory parallel computers. 
In an MPI parallel computation, 
the communication among distributed computer systems is 
handled by a communicator object such as \verb|MPI_COMM_WORLD|. 
Open Multi-Processing (OpenMP) is an application programming 
interface to control the multithreading computation 
on the shared-memory multiprocessor. 
The master thread 
forks several slave threads when an OpenMP directive 
such as ``\verb|#pragma omp parallel|''
appears in the program; and 
each thread concurrently executes the computation on the 
shared memory and finally joins the master thread 
at the end of the current block. 
The MPI and OpenMP are basically independent approaches to 
parallel computation. 
In recent years, hybrid parallel computing on massive supercomputers 
such as BlueGene/Q has become inevitable for obtaining 
a better computational performance. 
%

We develop a hybrid parallel C++ program using 
both MPI 
and 
OpenMP 
to calculate the four-point correlation functions 
of various $BB$ 
channels. 
The program works on 
general purpose computers such as 
the BlueGene/Q 
at the High Energy Accelerator Research 
Organisation 
(KEK) 
and 
HA-PACS at the University of Tsukuba. 
In a hybrid parallel computer program, 
the function 
\verb|MPI_Init_thread(int* argc, char ***argv, int required, int *provided)|
is called instead of 
\verb|MPI_Init(int* argc, char ***argv)|. 
For the third argument, 
we take the 
\verb|MPI_THREAD_MULTIPLE|
together with 
partitioning the \verb|MPI_COMM_WORLD| into a
number of sub-communicators in order 
to perform the 
multiple MPI communication through the sub-communicators 
concurrently from each forked multithreads.


Table~\ref{ElapsedTime} shows 
several elapsed times measured using 
the 32-node job class of BlueGene/Q at KEK 
during the calculations of the 
52 channels of the 
four-point correlation functions. 
The calculations are performed for 
a gauge configuration provided by CP-PACS and JLQCD Collaboration 
with a size of $L^3\times T = 16^3\times 32$~\cite{Ishikawa:2007nn}. 
Table~\ref{ElapsedTime} presents the results of 
the calculation of the four-point correlation functions and is 
divided into two parts:
the first part shows the data for the calculations of 
all single baryon blocks 
together with its FFT (step-1).
From the forms of the baryons' interpolating fields 
in Eqs.~(\ref{FieldOperator_N}), 
(\ref{FieldOperator_SLXi})$-$(\ref{FieldOperator_xxx}), 
it turns out that 
only six (constituents of) single baryon blocks, 
$B=p,\Sigma^{+}, \Xi^{0}, X_{u}, X_{d}$, and $X_{s}$, 
are actually computed so that 
all single baryon blocks, 
$B=p,n,\Sigma^{+}, \Sigma^{0}, \Sigma^{-}, \Xi^{0}, \Xi^{-}$, and $\Lambda$, 
are obtained from the above because of the symmetry 
under the interchange of the up and down quarks 
except for the overall phase factors 
in the isospin symmetric limit. 
The 
Second part shows the calculations of the 
52 four-point correlation functions 
$\sum_{\vec{X}}
 \left\langle 
   B_{1,\alpha}(\vec{X}+\vec{r},t)
   B_{2,\beta}(\vec{X},t)
   \overline{{\cal J}_{B_{3,\alpha^\prime} B_{4,\beta^\prime}}(0)}
 \right\rangle$, ($t=0,\cdots,T-1$)
from the baryon blocks by performing 
the summations of the indices of colour and spinor
together with its inverse FFT (step-2). 
The elapsed time is measured for various combinations of 
the number of MPI processes (\verb|tasks_per_node|) and 
the number of threads (\verb|OMP_NUM_THREADS|).
The elapsed time 
indicated by ``$64\times 1$'' is obtained from the so-called flat-MPI calculation. 
Sometimes, 
during hybrid parallel computations, there is a problem that 
hybrid parallel executions are not faster than the flat-MPI calculation. 
Our calculations do not show such a behaviour and the present program 
exhibits almost stable and reasonable performances for various combinations 
of the number of MPI processes and the number of threads. 

In step-1, the memory size can be reduced by sharing the 
memory of each baryon block if the same diagram appears 
(i.e. the components are numerically equivalent) 
throughout the 52 channels of the $BB$ four-point correlation functions. 
At present, this provides a benefit only for the memory usage, 
because computational cost of mapping the sharing of baryon blocks 
nullifies the gain in timing performance 
(see \ref{AGGREGATIONS} for further details). 
%
\begin{table}[t]
 \begin{minipage}{\textwidth}
  \begin{center}
   \caption{
Measured elapsed time for various hybrid parallel computation
of 
the 52 four-point correlation functions 
$\sum_{\vec{X}}
 \left\langle 
   B_{1,\alpha}(\vec{X}+\vec{r},t)
   B_{2,\beta}(\vec{X},t)
   \overline{{\cal J}_{B_{3,\alpha^\prime} B_{4,\beta^\prime}}(0)}
 \right\rangle$, ($t=0,\cdots,T-1$), 
by using the 32 node 
of BlueGene/Q on a $L^3\times T=16^3\times 32$ 
lattice, 
changing 
the number of MPI processes (tasks\_per\_node) and 
the number of threads (OMP\_NUM\_THREADS). 
A computational job consists of two steps; 
to calculate all of the single baryon blocks 
$[B_{\alpha}^{(I)}]$ 
together with its FFT (step-1), and 
to calculate the 52 four-point correlation functions 
by performing the summations of indices of colour and spinor 
together with its inverse FFT (step-2). 
%
\label{ElapsedTime}}
  \footnotesize
   \begin{tabular}{clllllll}
    \hline
{\scriptsize{\shortstack{ [tasks\_per\_node]~\qquad~\qquad~\\$\times$\tiny{[OMP\_NUM\_THREADS]}}}} &
$64\times 1$ & 
$32\times 2$ & 
$16\times 4$ & 
$8\times 4$ & 
$4\times 8$ & 
$2\times 16$ & 
$1\times 32$ \\
\hline
Step-1 & 00:14 & 00:16 & 00:09 & 00:09 & 00:07 & 00:06 & 00:06 \\
Step-2 & 00:10 & 00:11 & 00:12 & 00:12 & 00:12 & 00:13 & 00:14 \\
    \hline
   \end{tabular}
  \end{center}
 \end{minipage}
\end{table}

\section{Benchmark with the unified contraction algorithm\label{BENCHMARK}}

\begin{figure}[b]
 \begin{minipage}{0.49\textwidth}
  \includegraphics[width=0.97\textwidth]{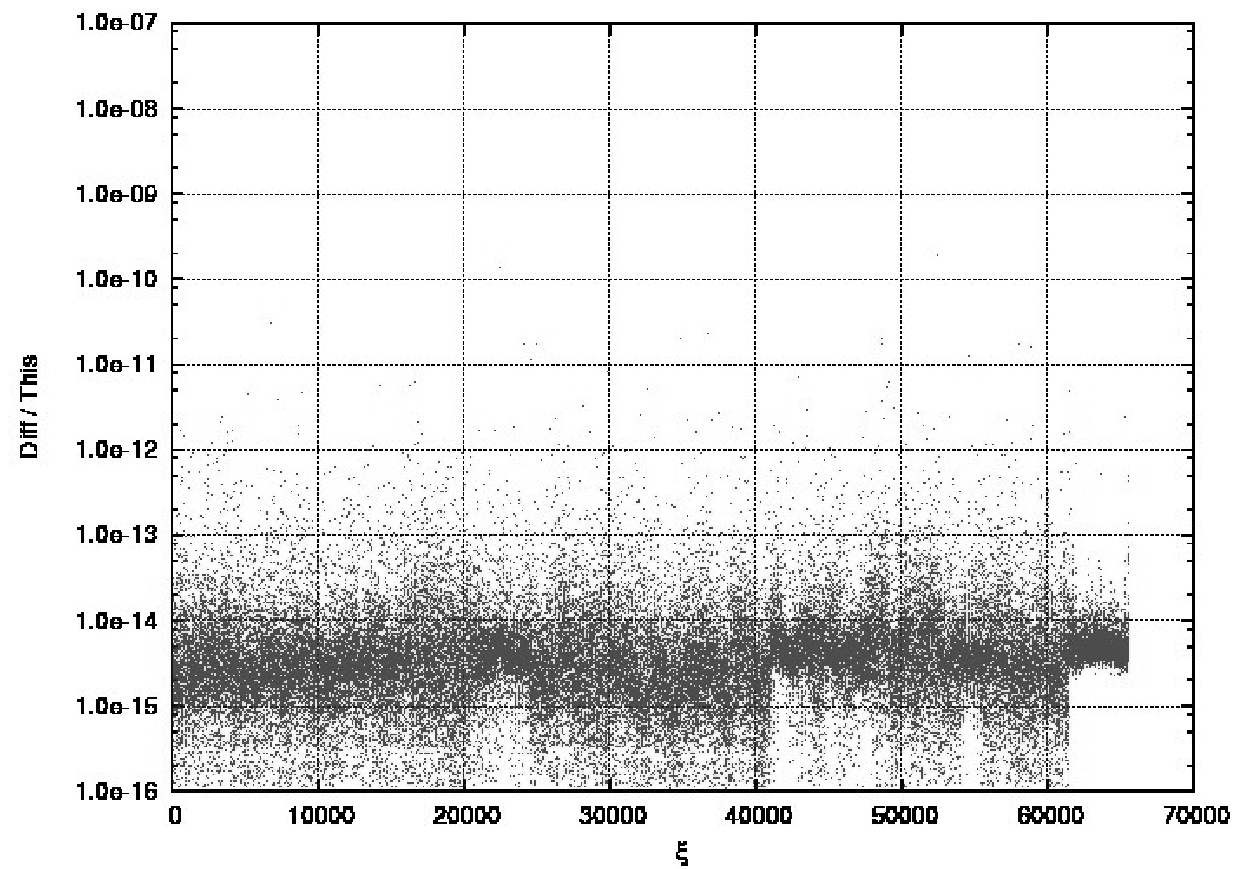}
 \end{minipage}~\hfill~
 \begin{minipage}{0.49\textwidth}
  \includegraphics[width=0.97\textwidth]{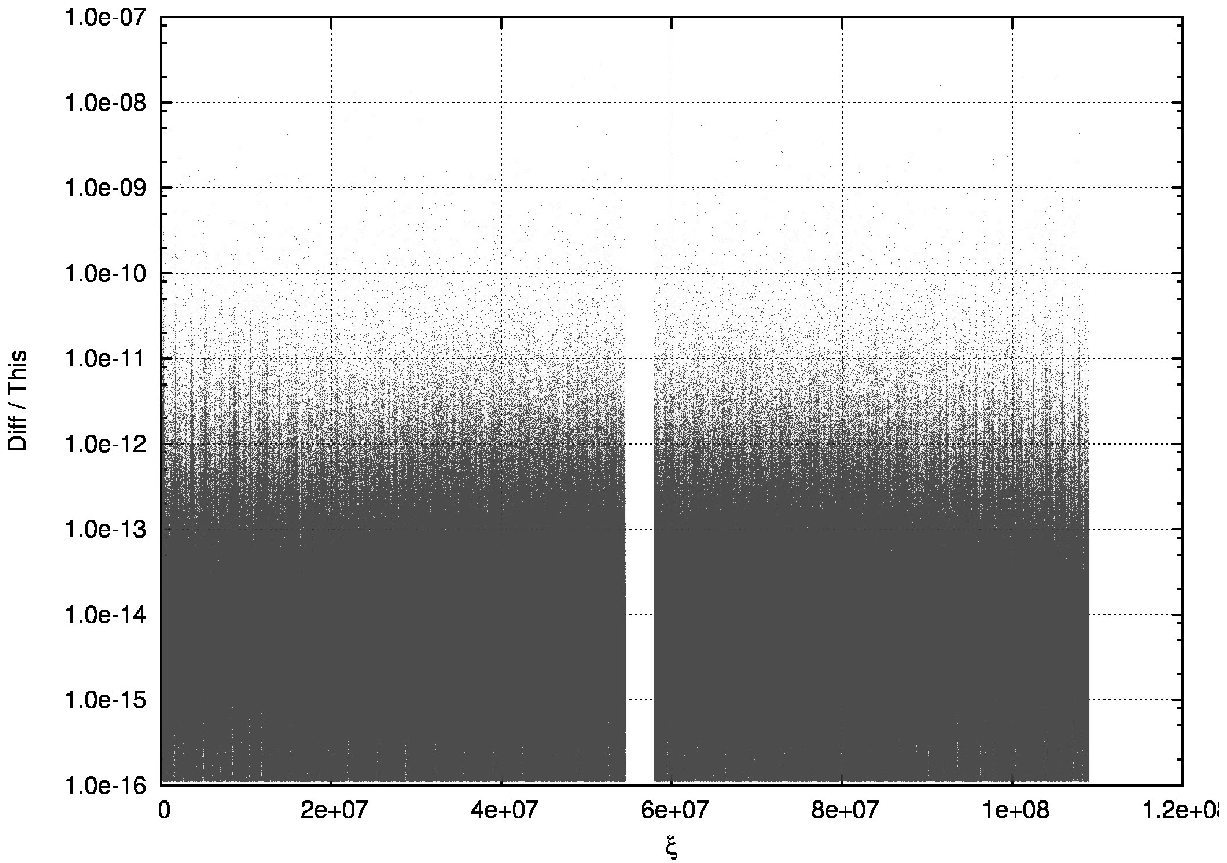}
 \end{minipage}
 \caption{
(Left)~The 
relative difference, $|\frac{\rm Diff}{\rm This\ work}|$, 
of the correlation function 
$\sum_{\vec{X}}
 \left\langle 
   p_{\alpha}(\vec{X}+\vec{r},t)
   \Lambda_{\beta}(\vec{X},t)
   \overline{{\cal J}_{p_{\alpha^\prime} \Lambda_{\beta^\prime}}(t_{0})}
 \right\rangle$ 
at $t-t_0=10$, 
between 
the effective block algorithm 
and 
the unified contraction algorithm 
as a function of 
one-dimensionally aligned data point 
$\xi = \tilde{\alpha} + 2(\tilde{\beta} + 2(\tilde{\alpha^\prime} +2(\tilde{\beta^\prime} +2(
x+16(y+16(z))))))$. 
(Right)~The 
relative difference 
of the correlators of 
entire 52 channels from $NN$ to $\Xi\Xi$ given in 
Eqs.~(\ref{GeneralBB_NN})$-$(\ref{GeneralBB_XX}), 
over 31 time-slices, 
$16^3$ points for spatial, and $2^4$ points for the spin degrees of 
freedom, 
between the effective block algorithm 
and 
the unified contraction algorithm 
as a function of one-dimensionally aligned data point 
$\xi = \tilde{\alpha} + 2(\tilde{\beta} + 2(\tilde{\alpha^\prime} +2(\tilde{\beta^\prime} +2(
x+16(y+16(z+16(c+52(
(t-t_0+T) \bmod T
))))))))$, where $c=0,\cdots,51$ selects 
one of the 52 channels 
provided that the correlator has non-vanishing value.
 \label{Fig_BENCHMARK} }
\end{figure}

In order to see the correctness of the present implementation of 
the effective block algorithm 
developed in Sec.~\ref{IMPLEMENTATION}, 
we benchmarked the numerical output with the corresponding data 
from the unified contraction algorithm~\cite{Doi:2012xd}. 
The benchmark has been done by using a gauge configuration 
provided from CP-PACS and JLQCD Collaboration 
with a size of $L^3\times T = 16^3\times 32$~\cite{Ishikawa:2007nn}. 
We have used a wall quark source with Coulomb gauge fixing and 
the periodic (Dirichlet) boundary condition has been imposed in the 
spatial (temporal) direction. 
Table~\ref{TABLE_BENCHMARK} shows just 16 lines of the 
comparisons as an example. 
For the correlator 
$R_{\alpha\beta\alpha^\prime\beta^\prime}(\vec{r}, t-t_{0})$
in the low-energy states, 
we 
adopt the Dirac representation and 
calculate upper (lower) two components of each spinor index 
to see the positive (negative) parity states of 
each single baryon (antibaryon) 
in the forward (backward) direction in time. 
Because of equivalence between 
the baryon-baryon states in forward direction in time 
and 
the antibaryon-antibaryon states in backward direction in time 
under the 
charge conjugation, parity, and time reversal 
operations, 
we effectively double our Monte Carlo samples by taking 
the data in both the forward and backward directions in time. 
We 
then 
reallocate the spinor indices, 
from 
$(\alpha,\beta,\alpha^\prime,\beta^\prime)$
to 
$(\tilde{\alpha},\tilde{\beta},\tilde{\alpha^\prime},\tilde{\beta^\prime})$, 
to run $0$ to $1$ for both cases 
in the numerical computation. 
The left panel of Figure~\ref{Fig_BENCHMARK} shows the 
relative difference, $|\frac{\rm Diff}{\rm This\ work}|$, 
of the correlator 
$\sum_{\vec{X}}
 \left\langle 
   p_{\alpha}(\vec{X}+\vec{r},t)
   \Lambda_{\beta}(\vec{X},t)
   \overline{{\cal J}_{p_{\alpha^\prime} \Lambda_{\beta^\prime}}(t_{0})}
 \right\rangle$ 
at $t-t_0=10$, 
between 
this effective block algorithm 
and 
the unified contraction algorithm 
as a function of 
one-dimensionally aligned data point 
$\xi = \tilde{\alpha} + 2(\tilde{\beta} + 2(\tilde{\alpha^\prime} +2(\tilde{\beta^\prime} +2(
x+16(y+16(z))))))$; 
there are $16^3 \times 2^4 = 65,536$ data points per time-slice per channel. 
The comparison is performed for all 52 channels over 31 time-slices, 
$16^3$ points for spatial, and $2^4$ points for the spin degrees of 
freedom. 
The right panel of Fig.~\ref{Fig_BENCHMARK} shows the 
result of the 
entire comparison between 
the effective block algorithm 
and 
the unified contraction algorithm, 
as a function of 
one-dimensionally aligned data point 
$\xi = \tilde{\alpha} + 2(\tilde{\beta} + 2(
\tilde{\alpha^\prime} +2(\tilde{\beta^\prime} +2(
x+16(y+16(z+16(c+52(
(t-t_0+T) \bmod T
))))))))$, where $c=0,\cdots,51$ selects 
one of 52 channels given in 
Eqs.~(\ref{GeneralBB_NN})$-$(\ref{GeneralBB_XX})\footnote{
The correlator, 
$\sum_{\vec{x}}\langle B_{1,\alpha}(\vec{x}+\vec{r})B_{2,\beta}(\vec{x})
\overline{B_{3,\alpha^{\prime}}B_{4,\beta^{\prime}}}\rangle$, 
vanishes 
due to the anticommutation relation of the baryon fields 
when two baryon fields become identical. 
It occurs in the following cases,
(i)~for the identical two baryons in the sink, 
$B_{1,\alpha} = B_{2,\beta}$, 
the correlator vanishes 
at $\vec{r}=$~a cyclic permutation of 
$(0,0,0),(L/2,0,0),(L/2,L/2,0)$, or $(L/2,L/2,L/2)$ 
under spatially periodic boundary conditions, 
(ii)~for the identical two baryons in the source, 
$\overline{B}_{3,\alpha^{\prime}}=\overline{B}_{4,\beta^{\prime}}$, 
the correlator vanishes 
under the present choice of wall quark source fields. 
These vanishing data points are not included in the figure. 
}. 
All numerical results are in good agreement 
with an accuracy of almost the double precision. 
%
\begin{table}[t]
 \begin{minipage}{\textwidth}
  \begin{center}
   \caption{
    Comparisons of numerical results between this work and 
    the other~\cite{Doi:2012xd} 
    are shown for only 16 lines of the four-point correlation function 
$\sum_{\vec{X}}
 \left\langle 
   p_{\alpha}(\vec{X}+\vec{r},t)
   \Lambda_{\beta}(\vec{X},t)
   \overline{{\cal J}_{p_{\alpha^\prime} \Lambda_{\beta^\prime}}(t_{0})}
 \right\rangle$ 
    at $t-t_0=10$. 
    ``Diff'' is the difference between ``This work'' and ``Other''.
    \label{TABLE_BENCHMARK}
   }
  \footnotesize
   \begin{tabular}{cccccccrrr}
    \hline
    $\tilde{\alpha}$ & $\tilde{\beta}$ & 
    $\tilde{\alpha^\prime}$ & $\tilde{\beta^\prime}$ & $x$ & $y$ & $z$ & 
    This work~\qquad~\qquad & 
    Other~\cite{Doi:2012xd}~\qquad~\qquad & 
    Diff~\quad~\quad \\
    \hline
  0 & 1 & 0 & 1 &   0 & 0 & 0 &  {\tt -3.075847140449e-21} &  {\tt -3.075847140449e-21} &  {\tt  3.4e-36} \\ 
  0 & 1 & 0 & 1 &   1 & 0 & 0 &  {\tt -8.786230541230e-21} &  {\tt -8.786230541230e-21} &  {\tt -3.0e-35} \\ 
  0 & 1 & 0 & 1 &   2 & 0 & 0 &  {\tt -1.138496114849e-20} &  {\tt -1.138496114849e-20} &  {\tt -3.8e-35} \\ 
  0 & 1 & 0 & 1 &   3 & 0 & 0 &  {\tt -8.109792412599e-21} &  {\tt -8.109792412599e-21} &  {\tt -2.4e-35} \\ 
  0 & 1 & 0 & 1 &   4 & 0 & 0 &  {\tt -1.086965914839e-20} &  {\tt -1.086965914839e-20} &  {\tt -2.9e-35} \\ 
  0 & 1 & 0 & 1 &   5 & 0 & 0 &  {\tt -9.926801964792e-21} &  {\tt -9.926801964792e-21} &  {\tt -6.0e-36} \\ 
  0 & 1 & 0 & 1 &   6 & 0 & 0 &  {\tt -6.647331180826e-21} &  {\tt -6.647331180826e-21} &  {\tt  2.0e-35} \\ 
  0 & 1 & 0 & 1 &   7 & 0 & 0 &  {\tt -1.640062750340e-21} &  {\tt -1.640062750340e-21} &  {\tt  5.0e-35} \\ 
  0 & 1 & 0 & 1 &   8 & 0 & 0 &  {\tt -2.553910496200e-21} &  {\tt -2.553910496200e-21} &  {\tt  7.0e-35} \\ 
  0 & 1 & 0 & 1 &   9 & 0 & 0 &  {\tt -1.250692150908e-22} &  {\tt -1.250692150907e-22} &  {\tt  7.3e-35} \\ 
  0 & 1 & 0 & 1 &  10 & 0 & 0 &  {\tt  4.866793580424e-21} &  {\tt  4.866793580424e-21} &  {\tt  9.4e-35} \\ 
  0 & 1 & 0 & 1 &  11 & 0 & 0 &  {\tt  1.379986127982e-20} &  {\tt  1.379986127982e-20} &  {\tt  1.2e-34} \\
  0 & 1 & 0 & 1 &  12 & 0 & 0 &  {\tt  1.680166855166e-20} &  {\tt  1.680166855166e-20} &  {\tt  9.9e-35} \\ 
  0 & 1 & 0 & 1 &  13 & 0 & 0 &  {\tt  1.176203581648e-20} &  {\tt  1.176203581648e-20} &  {\tt  6.2e-35} \\ 
  0 & 1 & 0 & 1 &  14 & 0 & 0 &  {\tt  2.994087733578e-21} &  {\tt  2.994087733579e-21} &  {\tt  3.1e-35} \\ 
  0 & 1 & 0 & 1 &  15 & 0 & 0 &  {\tt -9.904925605073e-22} &  {\tt -9.904925605073e-22} &  {\tt  2.4e-35} \\ 
    \hline
   \end{tabular}
  \end{center}
 \end{minipage}
\end{table}
%

\section{Summary\label{SUMMARY}}

In this paper, we present an approach for the efficient simultaneous 
calculation of a large number of four-point correlation functions, 
which are the primary quantities for studying the 
nuclear and hyperonic nuclear forces from lattice QCD. 
The effective block algorithm 
significantly reduces the number of iterations 
required for the Wick contraction, and 
is applied to calculate 
the 52 channels of four-point correlation functions 
in order to study the complete set of 
$BB$ 
interactions 
in the isospin symmetric limit. 
The elapsed time is measured for hybrid parallel computation 
on the BlueGene/Q supercomputer.  
The hybrid parallel executions of the $16^3\times 32$ lattice 
show reasonable performances for various combinations of 
the number of MPI processes and the number of threads.
The numerical values of the calculated 52 four-point correlation functions 
are compared with the results obtained using the unified contraction algorithm. 
We find that 
all 
numerical results 
are in good agreement and 
the two different algorithms give virtually identical results. 
This is advantageous for performing the 
large scale computation 
of various $BB$ potentials at the physical quark mass point.
%


\bigskip 

{%
The author would 
like to thank  
CP-PACS/JLQCD collaborations 
and 
ILDG/JLDG~\cite{ILDGJLDG} 
for 
allowing us to access the full QCD gauge configurations, 
and developers of Bridge++~\cite{BRIDGEPLUSPLUS}, 
and Dr.~T.~Doi for providing the numerical results of $52$ channels 
of NBS wave functions from the unified contraction algorithm. 
}
The author also thank maintainers of \verb|CPS++|~\cite{CPSPLUSPLUS}. 
Calculations in this paper have been performed 
by using the Blue Gene/Q computer 
under the ``Large scale simulation program'' 
at KEK (Nos. 12-11, 12/13-19). 
Part of this research was supported by Interdisciplinary 
Computational Science Program in CCS, University of Tsukuba. 
This research was supported in part
by Strategic Program for Innovative Research (SPIRE),
the MEXT Grant-in-Aid,  
Scientific Research on Innovative Areas and (C)
(Nos. 25105505, 16K05340).


\appendix


\section{The aggregation of effective blocks
\label{AGGREGATIONS}}

When calculating a large number of four-point correlation functions 
such as 52 channels of NBS wave functions 
simultaneously, 
we can economise on computer resource by 
aggregating 
the same effective blocks which appear several times through the 
whole calculation. 
In this section, 
we show how the aggregations are performed by 
considering the explicit form of the 
$\langle \Sigma^{+} n \overline{ \Sigma^{+} n }\rangle$ correlator. 

\subsection{Explicit form of the 
$\langle \Sigma^{+} n \overline{ \Sigma^{+} n }\rangle$ correlator}

\begin{figure}[t]
 \centering \leavevmode 
 \includegraphics[width=0.67\textwidth]{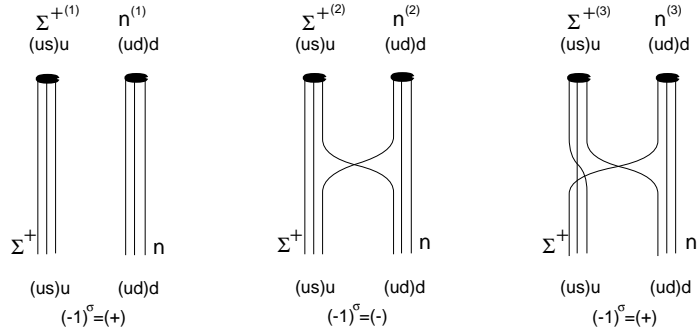}
 \caption{Diagrammatic representation of the four-point correlation function 
   $\langle \Sigma^{+} n \overline{ \Sigma^{+} n} \rangle$. 
   Three diagrams correspond to the three terms in 
   Eq.~(\ref{Qexchdiag_SpnnSp}). 
   The parity of each permutation is also shown as $(-1)^{\sigma}$. 
 \label{Fig_FFT_Spn} }
\end{figure}

The result of diagrammatic classification of the 
$\langle \Sigma^{+} n \overline{ \Sigma^{+} n }\rangle$ correlator 
is found in Table~\ref{NumberOfIterationsS01}. 
We show the explicit forms of the baryon blocks in this channel. 
The four-point correlator is given by 
%
\begin{equation}
 \begin{array}{l}
 \sum_{\vec{X}}
 \left\langle 0 \left|
  ~\Sigma^+_\alpha(\vec{X}+\vec{r},t)
  n_\beta(\vec{X},t)
  \overline{{\cal J}_{\Sigma^{+}_{\alpha^\prime} n_{\beta^\prime}}(t_0)}
 \right| 0 \right\rangle 
 \\
 \!\!\!\!\!\!=
  \sum_{\vec{X}}
  \varepsilon(1,6,2)
  \varepsilon(3,4,5)
  \varepsilon(1^\prime,6^\prime,2^\prime)
  \varepsilon(3^\prime,4^\prime,5^\prime)
  \\
  \times
  (C\gamma_5)(1,6) \delta(\alpha,2)
  (C\gamma_5)(3,4) \delta(\beta,5)
  (C\gamma_5)(1^\prime,6^\prime) \delta(\alpha^\prime,2^\prime)
  (C\gamma_5)(3^\prime,4^\prime) \delta(\beta^\prime,5^\prime)
 \\
 \times
  \langle
   u(1) s(6) u(2)
   u(3) d(4) d(5)
   \bar{d}(5^\prime) \bar{d}(4^\prime) \bar{u}(3^\prime)
   \bar{u}(2^\prime) \bar{s}(6^\prime) \bar{u}(1^\prime)
  \rangle,
 \end{array}
  \label{Spn.SpnnSp.usuudd}
\end{equation}
%
where
\begin{equation}
\vec{x}_1 = \vec{x}_2 = \vec{x}_6 = \vec{X} + \vec{r}, 
\qquad 
\vec{x}_3 = \vec{x}_4 = \vec{x}_5 = \vec{X}. 
\end{equation}
We have suppressed the explicit summations for the indices of 
colour and spinor in the right hand side. 
The last line in 
Eq.~(\ref{Spn.SpnnSp.usuudd}) 
is Wick contracted and 
represented in terms of the quark propagators, 
\begin{eqnarray}
 &&
  \langle
  u(1) s(6) u(2)
  u(3) d(4) d(5)
  \bar{d}(5^\prime) \bar{d}(4^\prime) \bar{u}(3^\prime)
  \bar{u}(2^\prime) \bar{s}(6^\prime) \bar{u}(1^\prime)
  \rangle
  \nonumber 
  \\
 &=&
 \left\{
  { \langle u(3) \bar{u}(3^\prime) \rangle }
  \det\left|
       \begin{array}{cc}
        \langle u(1) \bar{u}(1^\prime) \rangle &
	\langle u(1) \bar{u}(2^\prime) \rangle 
	\\
        \langle u(2) \bar{u}(1^\prime) \rangle &
	\langle u(2) \bar{u}(2^\prime) \rangle 
       \end{array}
      \right|
  \right.
 \nonumber 
 \\
 && \qquad
  \left.
   -
   { \langle u(3) \bar{u}(2^\prime) \rangle }
   \det\left|
	\begin{array}{cc}
	 \langle u(1) \bar{u}(1^\prime) \rangle &
	 \langle u(1) \bar{u}(3^\prime) \rangle
	 \\
	 \langle u(2) \bar{u}(1^\prime) \rangle &
	  \langle u(2) \bar{u}(3^\prime) \rangle
	\end{array}
       \right|
   \right.
  \nonumber 
  \\
 && \qquad
  \left.
   +
   { \langle u(3) \bar{u}(1^\prime) \rangle}
   \det\left|
	\begin{array}{cc}
	 \langle u(1) \bar{u}(2^\prime) \rangle &
	 \langle u(1) \bar{u}(3^\prime) \rangle
	 \\
	 \langle u(2) \bar{u}(2^\prime) \rangle &
	 \langle u(2) \bar{u}(3^\prime) \rangle
	\end{array}
       \right|
  \right\}
  \nonumber 
  \\
 && \times
  \det\left|
       \begin{array}{cc}
	\langle d(4) \bar{d}(4^\prime) \rangle &
        \langle d(4) \bar{d}(5^\prime) \rangle
	\\
	\langle d(5) \bar{d}(4^\prime) \rangle &
	\langle d(5) \bar{d}(5^\prime) \rangle
       \end{array}
      \right|
  \langle s(6) \bar{s}(6^\prime) \rangle. 
\label{Qexchdiag_SpnnSp}
\end{eqnarray}
Fig.~\ref{Fig_FFT_Spn} shows the diagrammatic representation of the 
correlator $\langle \Sigma^{+}n \overline{\Sigma^{+}n} \rangle$. 
%
%
The four-point correlation function is calculated using the FFT. 
We show the explicit forms of the three baryon-block pairs 
$\left([{\Sigma_{\alpha}^{+}}^{(1)}]\times[n_{\beta}^{(1)}]\right)$, \\
$\left([{\Sigma_{\alpha}^{+}}^{(2)}]\times[n_{\beta}^{(2)}]\right)$, 
$\left([{\Sigma_{\alpha}^{+}}^{(3)}]\times[n_{\beta}^{(3)}]\right)$. 

\bigskip \noindent
%
(i) {\bf ${\bfi{{\Sigma^{+}_{\alpha}}^{(1)}}}$ 
and $\bfi{n_{\beta}^{(1)}}$}

This is a product of two two-point correlators. 
\begin{equation}
 {R}_{\alpha\beta\alpha^\prime\beta^\prime}^{(1)}(\vec{r})
  \!=\! \sum_{\vec{X}} \left( [{{{\Sigma^{+}_{\alpha}}^{(1)}}}](\vec{X}\!+\!\vec{r}) 
  \!\times\! 
  [{n_{\beta}^{(1)}}](\vec{X})\right)_{\alpha^\prime\beta^\prime}
 \!=\! \sum_{\vec{X}} 
   [{{{\Sigma^{+}_{\alpha\alpha^\prime}}^{\!\!\!\!(1)}}}](\vec{X}+\vec{r}) 
   [{n_{\beta\beta^\prime}^{(1)}}](\vec{X}),
\end{equation}
where 
\begin{eqnarray}
 [{\Sigma^{+}_{\alpha\alpha^{\prime}}}^{(1)}](\vec{x})
  &=&
  \varepsilon(1,6,2) (C\gamma_5)(1,6) \delta(\alpha,2)
  \varepsilon(1^\prime,6^\prime,2^\prime)
  (C\gamma_5)(1^\prime,6^\prime) \delta(\alpha^\prime,2^\prime)
  \nonumber
  \\
 && \times 
  \det\left|
       \begin{array}{cc}
        \langle u(1) \bar{u}({ 1^\prime}) \rangle &
	\langle u(1) \bar{u}(2^\prime) \rangle 
	\\
        \langle u(2) \bar{u}(1^\prime) \rangle &
	\langle u(2) \bar{u}({ 2^\prime}) \rangle 
       \end{array}
      \right|
  \langle s(6) \bar{s}({ 6^\prime}) \rangle,
  \label{baryonblock_Sp1_SpnnSp}
  \\
 \left[{n}_{\beta\beta^{\prime}}^{(1)}\right](\vec{y})
 &=&
  \varepsilon(3,4,5)
  (C\gamma_5)(3,4) \delta(\beta,5)
  \varepsilon(3^\prime,4^\prime,5^\prime)
  (C\gamma_5)(3^\prime,4^\prime) \delta(\beta^\prime,5^\prime)
  \nonumber 
  \\
 && \times
  \langle u(3) \bar{u}({ 3^\prime}) \rangle
  \det\left|
       \begin{array}{cc}
        \langle d(4) \bar{d}({ 4^\prime}) \rangle &
	\langle d(4) \bar{d}(5^\prime) \rangle 
	\\
        \langle d(5) \bar{d}(4^\prime) \rangle &
	\langle d(5) \bar{d}({ 5^\prime}) \rangle 
       \end{array}
      \right|. 
  \label{baryonblock_n1_SpnnSp}
\end{eqnarray}
All of the summation of internal indices 
($\sum_{c_1, \cdots, c_6}$
$\sum_{c_1^\prime, \cdots, c_6^\prime}$
$\sum_{\alpha_1, \cdots, \alpha_6}$
$\sum_{\alpha_1^\prime, \cdots, \alpha_6^\prime}$)
can be performed 
separately for 
$[{\Sigma^{+}_{\alpha\alpha^{\prime}}}^{(1)}]$ and 
$\left[n_{\beta\beta^{\prime}}^{(1)}\right]$.

\bigskip \noindent
%
(ii) {\bf ${\bfi{{\Sigma^{+}_{\alpha}}^{(2)}}}$ 
and $\bfi{n_{\beta}^{(2)}}$}

This is an one-quark exchange diagram.
\begin{eqnarray}
 {R}_{\alpha\beta\alpha^\prime\beta^\prime}^{(2)}(\vec{r})
 &\!\!\!\!=&\!\!\!\! \sum_{\vec{X}} \left( 
  [{\Sigma^{+}}_{\alpha}^{(2)}](\vec{X}+\vec{r}) 
  \times
  [n_{\beta}^{(2)}](\vec{X}) \right)_{\alpha^\prime\beta^\prime}
 \nonumber
 \\
 &\!\!\!\!=&\!\!\!\! \sum_{\vec{X}}
  \!\!\sum_{c_2^\prime,c_3^\prime,
                                              \alpha_3^\prime}\!\!
  [{\Sigma^{+}_{\alpha}}^{(2)}](\vec{X}\!\!+\!\vec{r};
  { c_2^\prime,c_3^\prime,
                                          \alpha_3^\prime} )
  [n_{\beta\alpha^\prime\beta^\prime}^{(2)}](\vec{X};
  { c_2^\prime,c_3^\prime,
  \alpha_3^\prime} ), 
\end{eqnarray}
where
\begin{eqnarray}
 &&
  [{\Sigma^{+}_{\alpha}}^{(2)}]
  (\vec{x};c_2^\prime,c_3^\prime,
  \alpha_3^\prime)
  \nonumber
  \\
 &=&
  \varepsilon(1,6,2) (C\gamma_5)(1,6) \delta(\alpha,2)
  \varepsilon(1^\prime,6^\prime,2^\prime)
  (C\gamma_5)(1^\prime,6^\prime)
  \nonumber
  \\
 && \qquad \times
  \det\left|
        \begin{array}{cc}
	 \langle u(1) \bar{u}({ 1^\prime}) \rangle &
	 \langle u(1) \bar{u}(3^\prime) \rangle
	 \\
	 \langle u(2) \bar{u}(1^\prime) \rangle &
	  \langle u(2) \bar{u}({ 3^\prime}) \rangle
	\end{array}
      \right|
  \langle s(6) \bar{s}({ 6^\prime}) \rangle, 
  \label{baryonblock_Sp2_SpnnSp}
  \\
 &&
 [n_{\beta\alpha^\prime\beta^\prime}^{(2)}]
 (\vec{y};c_2^\prime,c_3^\prime,\alpha_3^\prime)
 \nonumber
 \\
 &=&
  \varepsilon(3,4,5)
  (C\gamma_5)(3,4) \delta(\beta,5)
  \varepsilon(3^\prime,4^\prime,5^\prime)
  (C\gamma_5)(3^\prime,4^\prime) \delta(\beta^\prime,5^\prime)
  \delta(\alpha^\prime,2^\prime)
  \nonumber
  \\
 && \qquad \times
  \langle u(3) \bar{u}({ 2^\prime}) \rangle
  \det\left|
       \begin{array}{cc}
        \langle d(4) \bar{d}({ 4^\prime}) \rangle &
	\langle d(4) \bar{d}(5^\prime) \rangle 
	\\
        \langle d(5) \bar{d}(4^\prime) \rangle &
	\langle d(5) \bar{d}({ 5^\prime}) \rangle 
       \end{array}
      \right|.
  \label{baryonblock_n2_SpnnSp}
\end{eqnarray}
We have additional arguments 
($c_2^\prime,c_3^\prime,
                                        \alpha_3^\prime$)
for the baryon blocks $[{\Sigma^{+}_{\alpha}}^{(2)}]$ and 
$[n_{\beta}^{(2)}]$ 
because of the exchange of $u$-quark in the source. 
Note that
the summation of 
$\alpha_2^\prime$ can be always omitted because of 
the presence of 
{
${\delta(\alpha^\prime,2^\prime)}$
located in $[n_{\beta}^{(2)}]$ 
}.

\bigskip \noindent
%
(iii) {\bf ${\bfi{{\Sigma^{+}_{\alpha}}^{(3)}}}$ 
and $\bfi{n_{\beta}^{(3)}}$}

This is another exchange diagram. 
\begin{eqnarray}
\!\!\!\! \!\!\!\! \!\!\!\!{R}_{\alpha\beta\alpha^\prime\beta^\prime}^{(3)}(\vec{r})
  &\!\!\!\!=&\!\!\!\!\!\! \sum_{\vec{X}} \left( [{\Sigma^{+}_{\alpha}}^{(3)}](\vec{X}+\vec{r})
  \times
  [n_{\beta}^{(3)}](\vec{X})\right)_{\alpha^\prime\beta^\prime}
 \nonumber 
 \\
  &\!\!\!\!=&\!\!\!\!\!\! \sum_{\vec{X}}
  \!\!\sum_{c_1^\prime,c_3^\prime,\alpha_1^\prime,\alpha_3^\prime}\!\!\!\!
  \!\![{\Sigma^{+}_{\alpha\alpha^\prime}}^{\!\!\!\!(3)}](\vec{X}\!\!+\!\vec{r};
  { c_1^\prime,c_3^\prime,\alpha_1^\prime,\alpha_3^\prime}  )
  [n_{\beta\beta^\prime}^{(3)}](\vec{X};
  { c_1^\prime,c_3^\prime,\alpha_1^\prime,\alpha_3^\prime}  ), 
\end{eqnarray}
where
\begin{eqnarray}
 &&
  [{\Sigma^{+}_{\alpha\alpha^\prime}}^{(3)}]
  (\vec{x};c_1^\prime,c_3^\prime,\alpha_1^\prime,\alpha_3^\prime)
  \nonumber 
  \\
 &=&
  \varepsilon(1,6,2) (C\gamma_5)(1,6) \delta(\alpha,2)
  \varepsilon(1^\prime,6^\prime,2^\prime)
  (C\gamma_5)(1^\prime,6^\prime) \delta(\alpha^\prime,2^\prime)
  \nonumber 
  \\
 && \qquad \times
  \det\left|
        \begin{array}{cc}
	 \langle u(1) \bar{u}({ 2^\prime}) \rangle &
	 \langle u(1) \bar{u}(3^\prime) \rangle
	 \\
	 \langle u(2) \bar{u}(2^\prime) \rangle &
	 \langle u(2) \bar{u}({ 3^\prime}) \rangle
	\end{array}
      \right|
  \langle s(6) \bar{s}({ 6^\prime}) \rangle, 
  \label{baryonblock_Sp3_SpnnSp}
  \\
 &&
 [n_{\beta\beta^\prime}^{(3)}]
 (\vec{y};c_1^\prime,c_3^\prime,\alpha_1^\prime,\alpha_3^\prime)
 \nonumber 
 \\
 &=&
  \varepsilon(3,4,5)
  (C\gamma_5)(3,4) \delta(\beta,5)
  \varepsilon(3^\prime,4^\prime,5^\prime)
  (C\gamma_5)(3^\prime,4^\prime) \delta(\beta^\prime,5^\prime)
  \nonumber 
  \\
 && \qquad \times
  \langle u(3) \bar{u}({ 1^\prime}) \rangle
  \det\left|
       \begin{array}{cc}
        \langle d(4) \bar{d}({ 4^\prime}) \rangle &
	\langle d(4) \bar{d}(5^\prime) \rangle 
	\\
        \langle d(5) \bar{d}(4^\prime) \rangle &
	\langle d(5) \bar{d}({ 5^\prime}) \rangle 
       \end{array}
      \right|. 
  \label{baryonblock_n3_SpnnSp}
\end{eqnarray}

\subsection{Finding reusable baryon blocks}

In the isospin symmetric limit, 
the single neutron correlator in Eq.~(\ref{baryonblock_n1_SpnnSp}) 
is identical with 
the single proton  correlator in Eq.~(\ref{baryonblock_p1_pLLp}) 
because 
the interpolating fields of proton and neutron 
in Eq.~(\ref{FieldOperator_N}) are 
symmetric under the interchange of the up and down quarks 
except for the overall phase factors. 
Thus we may avoid the actual numerical calculation of the 
$[n_{\beta\beta^\prime}^{(1)}](\vec{y})$
in $\langle~\Sigma^+ n~ \overline{\Sigma^+ n}~\rangle$ 
by using the result of 
$[p_{\alpha\alpha^\prime}^{(1)}](\vec{x})$
in $\langle p \Lambda \overline{p \Lambda} \rangle$
instead: 
\begin{equation}
[n_{\beta\beta^\prime}^{(1)}](\vec{y})_{
\langle~\Sigma^+ n~ \overline{\Sigma^+ n}~\rangle}
=
\left(
[p_{\alpha\alpha^\prime}^{(1)}](\vec{x})_{
\langle p \Lambda \overline{p \Lambda} \rangle}
\right)_{\scriptsize
\left(
 \begin{array}{c}
  \alpha \rightarrow \beta
   \\
  \alpha^\prime \rightarrow \beta^\prime
   \\
  \vec{x} \rightarrow \vec{y}
 \end{array}
\right)
}.
\label{aggregation_n1_SpnnSp}
\end{equation}
The usage of Eq.~(\ref{aggregation_n1_SpnnSp}) gives right result 
provided that 
the spatial reflection in momentum space is taken into account 
when performing the FFT with the replacement of 
the space coordinate $\vec{x} \rightarrow \vec{y}$. 
See Eq.~(\ref{LN.pLLp.FFT}), where the argument of the second 
baryon is $(-\vec{q})$ while the first baryon serves $(\vec{q})$. 
The above first example might be a very trivial case. 
The second example is to find that the 
$[n_{\beta}^{(2)}](\vec{y})$ 
in $\langle~\Sigma^+ n~ \overline{\Sigma^+ n}~\rangle$ 
in Eq.~(\ref{baryonblock_n2_SpnnSp}) 
is a special case of 
$[n_{\beta}^{(3)}](\vec{y})$ 
in $\langle~\Sigma^+ n~ \overline{\Sigma^+ n}~\rangle$ 
in Eq.~(\ref{baryonblock_n3_SpnnSp}), 
\begin{equation}
[n_{\beta\alpha^\prime\beta^\prime}^{(2)}]
 (\vec{y};c_2^\prime,c_3^\prime,
                                                \alpha_3^\prime)
_{\langle \Sigma^+ n \overline{\Sigma^+ n} \rangle}
=
\left(
[n_{\beta\beta^\prime}^{(3)}]
 (\vec{y};c_1^\prime,c_3^\prime,\alpha_1^\prime,\alpha_3^\prime)_{
\langle \Sigma^+ n \overline{\Sigma^+ n}\rangle}
\right)_{\scriptsize
\left(
\begin{array}{c}
c_1^\prime      \rightarrow c_2^\prime \\
\alpha_1^\prime \rightarrow \alpha_2^\prime=\alpha^\prime
\end{array}
\right)
}.
\end{equation}
These kinds of reusable baryon blocks can be found in various parts in the 
entire 52 channels of the NBS wave functions. 
We list only a few more examples 
that figuring in one's head is possible 
from the above explicit forms of the baryon blocks shown in this paper: 
\begin{equation}
\![n_{\beta\beta^\prime}^{(3)}]
 (\vec{y};c_1^\prime,c_3^\prime,\alpha_1^\prime,\alpha_3^\prime)_{
\langle \Sigma^+ n \overline{\Sigma^+ n} \rangle}
\!\!=\!\!
\left(
[p_{\alpha\alpha^\prime}^{(4)}]
 (\vec{x};c_4^\prime,c_5^\prime,\alpha_4^\prime,\alpha_5^\prime)_{
\langle p  \Lambda  \overline{p \Lambda} \rangle}
\right)_{\scriptsize
\!\!\!\!\left(
 \begin{array}{c}
  \!\!\!\!(c_4^\prime,\alpha_4^\prime) \rightarrow (c_3^\prime,\alpha_3^\prime) \\
  \!\!\!\!(c_5^\prime,\alpha_5^\prime) \rightarrow (c_1^\prime,\alpha_1^\prime) \\
  \alpha \rightarrow \beta \\
  \alpha^\prime \rightarrow \beta^\prime \\
  \vec{x} \rightarrow \vec{y}
 \end{array}
\!\!\!\!\right)
},
\end{equation}
\begin{equation}
 [p_{\alpha\alpha^\prime\beta^\prime}^{(6)}](\vec{x};
    c_2^\prime,c_6^\prime,
                               \alpha_6^\prime)
                                           _{
\langle p \Lambda \overline{ p \Lambda} \rangle}
=
  \left(
   [p_{\alpha\beta^\prime}^{(4)}](\vec{x};
       c_1^\prime,c_6^\prime,\alpha_1^\prime,\alpha_6^\prime)_{
                \langle p \Lambda \overline{ p \Lambda} \rangle}
  \right)_{\scriptsize
   \left(
    \begin{array}{c}
     c_1^\prime      \rightarrow c_2^\prime \\
     \alpha_1^\prime \rightarrow \alpha_2^\prime=\alpha^\prime
    \end{array}
   \right)
},
\end{equation}
\begin{equation}
 [\Lambda_{\beta\alpha^\prime
                                          }^{(2)}](\vec{y};
c_2^\prime,c_3^\prime
                                          )
                                           _{
\langle p \Lambda \overline{ p 
                               X_{u}
                                       } \rangle }
  = \left(
     [\Lambda_{\beta
                                 }^{(5)}](\vec{y};
c_1^\prime,c_3^\prime,\alpha_1^\prime
                                                     )_{
\langle p \Lambda \overline{ p 
                               X_{u}
                                       } \rangle 
}
    \right)_{\scriptsize
     \left( 
      \begin{array}{c}
       c_1^\prime      \rightarrow c_2^\prime \\
       \alpha_1^\prime \rightarrow \alpha_2^\prime=\alpha^\prime
      \end{array}
     \right)
  }.
\end{equation}
Table~\ref{MEMORY_USAGE} summarises that how the memory size reduces by 
considering the aggregations of the effective baryon blocks throughout 
the entire 52 channels of the NBS wave functions. 
%
\begin{table}[t]
 \begin{minipage}{\textwidth}
  \begin{center}
   \caption{
     The number of effective baryon block objects declared and 
     the memory size of each baryon block object 
     for a calculation of the 
     52 NBS wave functions given in 
     Eqs.~(\ref{GeneralBB_NN})$-$(\ref{GeneralBB_XX}) 
     for taking a normal approach or the improved (aggregative) approach 
     described in \ref{AGGREGATIONS}. 
     The ratio of aggregative to normal is also presented. 
     For the $2+1$ flavour lattice QCD calculation, 
     a quantity of $X_{d}$ can be replaced by the 
     corresponding quantity from $X_{u}$. 
     Thus no actual memory is required for the $X_{d}$ 
     for the improved algorithm. 
    \label{MEMORY_USAGE}
   }
  \footnotesize
   \begin{tabular}{llcccccc}
    \hline
    &       & $p$ & $\Sigma^{+}$ & $\Xi^{0}$ & $X_{u}$ & $X_{d}$ & $X_{s}$ \\
    \hline
    {\scriptsize{\shortstack{Number of \\baryon blocks}}} & {\scriptsize{Normal}} & 
    $304$  & $124$  & $298$  & $1784$ & $1784$ & $984$  \\
                & {\scriptsize{Aggregative}} & 
    $28$   & $18$   & $19$   & $36$   & $0$    & $32$   \\
           & {\scriptsize{Ratio~(\%)}}  & 
    $9.21$ & $14.5$ & $6.38$ & $2.02$ & $0$    & $3.25$ \\
    \hline
    {\scriptsize{\shortstack{Memory size\\~($\times 16$~Bytes/site)}}} & {\scriptsize{Normal}} & 
    $120144$ & $46960$ & $106104$ & $554896$ & $554896$ & $305896$ \\
    \quad~{\scriptsize{}} & {\scriptsize{Aggregative}} & 
    $10952$  & $7208$  & $7784$   & $12128$  & $0$      & $10968$  \\
            & {\scriptsize{Ratio~(\%)}} & 
    $9.12$   & $15.3$  & $7.34$   & $2.19$   & $0$      & $3.59$   \\
    \hline
   \end{tabular}
  \end{center}
 \end{minipage}
\end{table}







\end{document}